\documentclass[12pt,a4paper,oneside]{article}
\usepackage{graphicx,amssymb,amsmath,color}
\usepackage[colorlinks=true,urlcolor=blue,anchorcolor=blue,citecolor=red,filecolor=blue
,linkcolor=blue,menucolor=blue,linktocpage=true,pdfproducer=medialab,pdfa=true
]{hyperref}
\usepackage{cite}
\usepackage{relsize}
\usepackage{enumerate}
\usepackage[margin=2.3 cm]{geometry}
\usepackage{slashed}
\usepackage{multirow}
\usepackage[normalem]{ulem}
\usepackage[dvipsnames, x11names]{xcolor}
\usepackage[square,comma,sort&compress,numbers]{natbib}
\usepackage{soul}
\usepackage{bigints}
\usepackage{relsize,exscale}
\usepackage{stmaryrd}

\usepackage{cleveref}
\usepackage{cancel}

\def\beq{\begin{equation}}
\def\eeq{\end{equation}}
\def\bea{\begin{eqnarray}}
\def\eea{\end{eqnarray}}

\def\bit{\begin{itemize}}
\def\eit{\end{itemize}}

\def\l{\left}
\def\r{\right}

\def\baa{\begin{array}}
\def\eaa{\end{array}}

\def\simgt{\mathrel{\lower2.5pt\vbox{\lineskip=0pt\baselineskip=0pt
           \hbox{$>$}\hbox{$\sim$}}}}
\def\simlt{\mathrel{\lower2.5pt\vbox{\lineskip=0pt\baselineskip=0pt
           \hbox{$<$}\hbox{$\sim$}}}}

\def\P{\mathcal{P}}

\def\bfc{\begin{figure}\begin{center}}
\def\efc{\end{center}\end{figure}}
\def\nn{\nonumber\\}

\definecolor{chromeyellow}{rgb}{1.0, 0.65, 0.0}
\definecolor{darkcoral}{rgb}{0.8, 0.36, 0.27}
\definecolor{cadmiumgreen}{rgb}{0.0, 0.42, 0.24}

\begin{document}

\begin{flushright}
\hspace{3cm} 
SISSA  12/2024/FISI \\
DESY--24--080
\end{flushright}
\vspace{.6cm}
\begin{center}

\hspace{-0.4cm}{\Large \bf 
The hydrodynamics of inverse phase transitions\\
}

\vspace{1cm}{}
\end{center}

\begin{center}
Giulio Barni$^{*\,1,2}$, Simone Blasi,$^{\dagger\,3,4}$ and Miguel Vanvlasselaer$^{\ddagger \, 4}$ \\
\vskip0.4cm

{\hspace{-.08cm}\it $^1$ SISSA International School for Advanced Studies, Via Bonomea 265, 34136, Trieste, Italy}\par
\vskip0.2cm
{\it $^2$ INFN - Sezione di Trieste, Via Bonomea 265, 34136, Trieste, Italy}\par
\vskip0.2cm
{\it $^3$ Deutsches Elektronen-Synchrotron DESY, Notkestr.~85, 22607 Hamburg, Germany} \par
\vskip0.2cm
{\it $^4$ Theoretische Natuurkunde and IIHE/ELEM, Vrije Universiteit Brussel,
\& The International Solvay Institutes, Pleinlaan 2, B-1050 Brussels, Belgium }
\vskip1.cm
\end{center}

\bigskip \bigskip \bigskip

\begin{abstract}
First order phase transitions are violent phenomena that occur when the state of the universe evolves abruptly from one vacuum to another. A \emph{direct} phase transition connects a local vacuum to a deeper vacuum of the zero--temperature potential, and the energy difference between the two minima manifests itself in the acceleration of the bubble wall. In this sense, the transition is triggered by the release of vacuum energy. On the other hand, an \emph{inverse} phase transition connects a deeper minimum of the zero--temperature potential to a higher one, and the bubble actually expands against the vacuum energy. The transition is then triggered purely by thermal corrections. 
We study for the first time the hydrodynamics and the energy budget of inverse phase transitions. We find several modes of expansion for inverse bubbles, which are related to the known ones for direct transitions by a mirror symmetry. We finally investigate the friction exerted on the bubble wall and comment on the possibility of runaway walls in inverse phase transitions.

\end{abstract}

\vfill
\noindent\line(1,0){188}
{\footnotesize{ \\ 
\text{$^*$~gbarni@sissa.it}\\
\text{$^\dagger$~simone.blasi@desy.de}\\
\text{$^\ddagger$~miguel.vanvlasselaer@vub.be}}}

\newpage

\hrule
\tableofcontents
\vskip.8cm
\hrule

\section{Introduction}

Phase transitions (PTs) in the early universe plasma, usually called \emph{cosmological} phase transitions, have recently received much attention mostly due to the broad range of interesting consequences that they can bring to the early universe thermal history. From a phenomenological perspective, cosmological phase transitions can be at the origin of the baryogenesis~\cite{Kuzmin:1985mm, Shaposhnikov:1986jp,Nelson:1991ab,Carena:1996wj,Cline:2017jvp,Long:2017rdo,Bruggisser:2018mrt,Bruggisser:2018mus, Bruggisser:2022rdm,Morrissey:2012db,Azatov:2021irb, Huang:2022vkf, Baldes:2021vyz, Chun:2023ezg}, the production of heavy dark matter~\cite{Falkowski:2012fb, Baldes:2020kam,Hong:2020est, Azatov:2021ifm,Baldes:2021aph, Asadi:2021pwo, Lu:2022paj,Baldes:2022oev, Azatov:2022tii, Baldes:2023fsp,Kierkla:2022odc, Giudice:2024tcp}, primordial black holes~\cite{10.1143/PTP.68.1979,Kawana:2021tde,Jung:2021mku,Gouttenoire:2023naa,Lewicki:2023ioy} and possibly observable gravitational wave (GW)~\cite{Witten:1984rs,Hogan_GW_1986,Kosowsky:1992vn,Kosowsky:1992rz,Kamionkowski:1993fg}. Moreover, from a theoretical perspective, PTs between a local minimum and a deeper, local or global, minimum are commonplace in quantum field theory, where it is believed that the vacuum structure is a complicated manifold. In a related way, PTs appear naturally in a large variety of motivated BSM models like composite Higgs~\cite{Pasechnik:2023hwv, Azatov:2020nbe,Frandsen:2023vhu, Reichert:2022naa,Fujikura:2023fbi}, extended Higgs sectors~\cite{Delaunay:2007wb, Kurup:2017dzf, VonHarling:2017yew, Azatov:2019png, Ghosh:2020ipy,Aoki:2021oez,Badziak:2022ltm, Blasi:2022woz,Agrawal:2023cgp,Banerjee:2024qiu}, axion models~\cite{DelleRose:2019pgi, VonHarling:2019gme}, dark Yang-Mills sectors~\cite{Halverson:2020xpg,Morgante:2022zvc}, $B-L$ breaking sectors~\cite{Jinno:2016knw, Addazi:2023ftv}. 

\begin{figure}
 \centering  \includegraphics[width=0.48\textwidth]{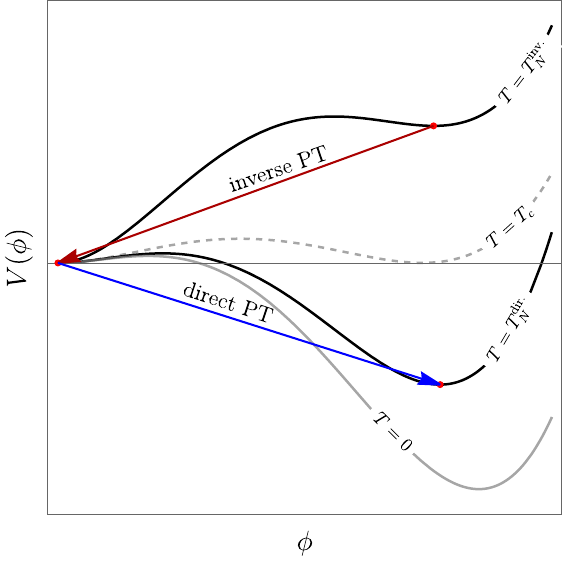}
 \caption{Schematic representation of the \emph{thermally corrected} potential of a phase transition triggered by the vacuum energy, denoted as direct PT (blue arrow), and a phase transition against the vacuum energy, triggered by thermal corrections, referred to as inverse phase transition (darker red arrow). {In fact, focusing only on the $T=0$ potential, the inverse PT can be seen to actually proceed from a point with lower vacuum energy
 to a point with higher vacuum energy. In this sense, the inverse PT (which is made possible by the thermal effects) goes against the $T=0$ vacuum energy.
 }}
 \label{fig:schematic}
 \end{figure}

For all these reasons, the hydrodynamics of cosmological phase transitions have been intensively studied in the past, alongside with their hydrodynamical properties, their efficiency to turn vacuum energy into bulk motion, sound speed effects~\cite{Espinosa:2010hh,Giese:2020rtr,Giese:2020znk,Wang:2021dwl,Ajmi:2022nmq,Tenkanen:2022tly,Wang:2022lyd,Wang:2023jto} and gravitational wave imprint~\cite{Caprini:2019egz, Hindmarsh:2015qta, Hindmarsh:2017gnf} (see for example \cite{Athron:2023xlk} for a review). A thorough classification of the difference modes of expansion of bubbles wall has been presented~\cite{Ignatius:1993qn, Laine:1993ey, Kurki-Suonio:1995rrv, Giombi:2023jqq}. Five consistent types of solutions survived the examination: weak and Chapman-Jouguet (CJ) deflagrations, weak and CJ detonations and hybrid solutions, which are supersonic deflagrations glued to rarefaction waves.  The collapse of cosmological droplets, because of their possible impact on the production of GW~\cite{PhysRevD.106.103524} and PBH production~\cite{Lewicki:2023mik}, also received attention. In a direct phase transition, the vacuum undergoes a transition from a local higher minimum of the zero--temperature potential to a deeper minimum, as presented by the blue arrow (direct PT) in Fig.\ref{fig:schematic}. The acceleration of the bubbles of the new phase is then triggered mostly by the vacuum energy release. 

A much less studied situation is the expansion of bubbles of inverse phase transitions, where the transition is from a lower minimum  (of the zero--temperature potential) to a higher one, as presented in solid darker red arrow (inverse PT) in Fig.\ref{fig:schematic}.  During inflation, we expect the true zero temperature vacuum to be populated and the symmetries to be broken by the vacuum expectation values of the scalar fields. The subsequent reheating will then increase the temperature of the bath and push the scalar fields to the origin, to the restoration of symmetries. The transition to the symmetric vacuum can then proceed via the nucleation of \emph{inverse} bubbles expanding \emph{against} the vacuum energy.
Such a phenomenon has been studied in~\cite{Buen-Abad:2023hex, Kolesova:2023mno} in the context of reheating after inflation,  during the superheated deconfinement PTs of QCD that could occur during neutron stars mergers~\cite{Casalderrey-Solana:2022rrn} and {in some fast walls realisations of baryogenesis \cite{Caprini:2011uz} }. 
{Nevertheless, let us stress that our analysis is more general and only assumes that the universe
finds itself in a false vacuum with lower ($T=0$) vacuum energy compared to the true vacuum. While reheating is the
most straightforward realization, this could as well be the outcome of a specific
thermal history during the standard cooling of the universe.}
These ``inverse'' bubbles will then differ in many respects from the bubbles nucleated during direct PTs.

In this paper, we present a thorough study of the modes of expansion of \emph{inverse} phase transitions and discuss their energy budget. We also investigate the velocity of such bubbles. Here is a summary of the main results of our study: 
\begin{itemize}
\item The main parameters controlling the expansion are the wall velocity $\xi_w$ and the strength of the PT $-\alpha_N\equiv |\epsilon|/\rho_N$, where $\epsilon$ is the difference of vacuum energy, and $\rho_N$ is the radiation energy density at the nucleation temperature $T_{N}$. 
\item We find five different types of consistent solutions: weak and Chapman-Jouguet (CJ) inverse deflagrations, weak and CJ inverse detonations, and inverse hybrid solutions. They however differ in several respects from their direct counterparts. In the plasma frame, the fluid velocities are typically negative (toward the inside of the bubble), meaning that the plasma is being sucked into the bubble.  
\item We calculate the efficiency of the energy transfer from the phase transition to the bulk flow motion, which acts as the source of gravitational waves.
\item We also study the pressure exerted on inverse bubbles. We discover a hydrodynamic obstruction to the expansion of inverse phase transition, very similar to the obstruction to the expansion of direct ones. In the regime of fast bubbles, where the collisionless approach can be applied, we find a new Boltzmann suppression of the plasma pressure.
\end{itemize}

The remainder of this article is organised as follows: in Section \ref{sec:direct_PT} we present a review of known results of the hydrodynamics of the direct phase transitions, in Section \ref{sec:InvPT}, we present the study of the hydrodynamics of \emph{inverse} phase transitions, in Section \ref{sec:energy_budget}, we discuss the energy budget of the inverse phase transition as well as the efficiency factors. In Section \ref{sec:friction_inv_PT}, we discuss the friction effects on the bubble wall and in Section \ref{sec:runaway_inv_PT}, the possibility of runaway solutions is discussed.  Finally, we conclude in Section \ref{sec:conclusion}. 

\section{Direct phase transitions: a reminder}
\label{sec:direct_PT}

In this section, we review the expansion modes of a cosmological bubble that forms during a first-order PT within a primordial plasma background while the universe cools down. This type of transition is the one considered in most cosmological applications, and it is characterized by a release of vacuum energy into the plasma. For this reason, we refer to it as \emph{direct}. The case of \emph{inverse} PTs against the vacuum energy will be the topic of Sec.\,\,\ref{sec:InvPT}.

\subsection{Matching across discontinuities}
\label{sec:matching_direct_PT}
The hydrodynamics of the coupled system, where a nucleated bubble expands within the primordial plasma, can be described by the conservation of the total energy-momentum tensor. The energy-momentum tensor contains two pieces: i) the scalar background, which generates the bubble wall profile that we denote $\phi$, and ii) the plasma that we denote $f$ and that we model as a perfect fluid. Those two contributions respectively read
\begin{subequations}
\begin{align}
T_{\phi}^{\mu\nu}&=(\partial^\mu\phi)\partial^\nu\phi-g^{\mu\nu}\left[\frac{1}{2}(\partial\phi)^2-V(\phi)\right]\ ,
\qquad \text{(scalar field component)}
\\
T_f^{\mu\nu}&=(e_f+p_f)u^\mu u^\nu- g^{\mu\nu}p_f \ , \qquad \text{(plasma component)}
\end{align}
\end{subequations}
where $u^\mu=\gamma(v)(1, \Vec{v})$ is the fluid four-velocity in the plasma frame with the Lorentz boost factor $\gamma(v)=1/\sqrt{1-v^2}$, $e_f$ and $p_f$ are the fluid energy density and pressure, that vanish at zero temperature. $V$ is the effective (loop-resummed) scalar potential. However, one usually combines the fluid energy density and pressure with the effective scalar potential energy, $e=e_f+V(\phi)$, $p=p_f-V(\phi)$. The advantage of writing the energy-momentum tensors in terms of $e$ and $p$ is that the matching conditions for hydrodynamic quantities take the standard form that appears commonly in the literature. Note that the fluid \emph{enthalpy} writes $w=e_f+p_f=e+p$. Therefore, in terms of $e$ and $p$, the energy-momentum tensor for the fluid then takes the following form
\begin{align}
    T_f^{\mu\nu}&=(e+p)u^\mu u^\nu- g^{\mu\nu}[p+V(\phi)]\ .
\end{align}
Then, the conservation of the energy-momentum tensor is given by 
\bea 
\label{eq: energy cons}
\nabla_\mu T^{\mu\nu}=\nabla_\mu \l( T_\phi^{\mu\nu}+T_f^{\mu \nu} \r)=0 \ .
\eea

\begin{figure}
    \centering
    \includegraphics[width=.5\textwidth]{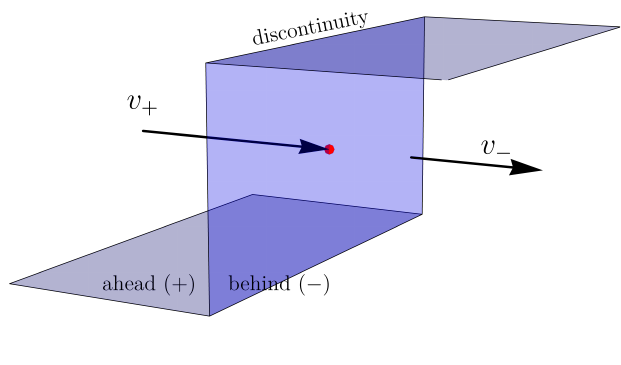}
 \caption{Pictorial representation of a discontinuity interface in the wall frame. The fluid ahead of the discontinuity, (+), is coming towards the wall in the region behind, $(-)$. For direct PT $(+)$ is (generically) the symmetric phase, while $(-)$ is the broken one.}
    \label{fig:sketch}
\end{figure}

Hydrodynamical flows can develop discontinuities such as shock waves and reaction fronts, across which the bulk quantities undergo a jump, as pictorially presented in Fig.\ref{fig:sketch}. The conservation equations in Eq.\eqref{eq: energy cons} can then be used to derive junction conditions of these quantities. Those will serve as boundary conditions for the smooth evolution of the fluid on both sides of the discontinuity. By integrating Eq. \eqref{eq: energy cons} over a volume containing the interface and using Stokes' theorem, we arrive at the continuity equations governing the flow of energy-momentum
\bea 
(T_+^{z\nu}-T_-^{z\nu})n_\nu=0 \ ,\qquad (T_+^{t\nu}-T_-^{t\nu})n_\nu=0 \ , 
\eea 
where $n_\mu=(0,0,0,1)$ is the unit 4-vector perpendicular to the bubble interface. Under the assumption that the flux along the 3-direction is $u^{\mu}=\gamma(z) (1,0,0,-v(z))$, one can obtain the junction conditions
\begin{subequations}
\label{eq:junctionAB}
\begin{align}
    w_+\gamma_+^2v_+ &=w_-\gamma_-^2v_-\ ,\label{eq:conditionA}\\
     w_+\gamma_+^2v_+^2+p_+ &=w_-\gamma_-^2v_-^2+p_-\ ,
   \label{eq:conditionB}
\end{align}
\end{subequations}
where $w\equiv e+p$ is the enthalpy and where the subscript ``$\pm$'' denotes quantities in front of/behind the bubble wall, so that $-$ always represents the interior of the bubble. To be explicit, $w_+=w_s(T_+)$, $w_-=w_b(T_-)$ (and similarly for $p_{\pm }$), where the label ``$s/b$'' denotes the symmetric/broken phase, see Fig.~\ref{fig:sketch}. 
Upon rearranging the junction conditions, we arrive at the familiar relations between the velocities, the energies and the pressures,
\bea 
\label{eq:rel_velocities}
v_+v_- = \frac{p_+- p_-}{e_+-e_-}\ ,\qquad \frac{v_+}{v_-} = \frac{e_-+p_+}{e_++p_-}\ .
\eea 
We remind that the velocities $v_\pm$ have to be understood in the \emph{front frame} (the frame where the discontinuity is at rest). 
To advance and determine the solutions for the system of hydrodynamical equations, we must assume a specific equation of state (EoS) for the plasma. This EoS will represent a function that relates various thermodynamic quantities.

\subsection{Introducing an Equation of State}
To make further progress, we need to introduce an Equation of State (EoS), which for simplicity we take to be the bag EoS:
    \begin{align}
\label{eq:bag_eos}
    &e_+(T)=a_+  T^4+\epsilon_+,\qquad p_+(T)=\frac{1}{3}a_+ T^4-\epsilon_+,
    \nonumber
    \\
    &e_-(T)=a_- T^4 +\epsilon_- , \qquad  p_-(T)=\frac{1}{3}a_- T^4- \epsilon_-,
\end{align}
where $a_\pm$ and $\epsilon_{\pm}$ are constants and we used the convention $\epsilon_+ - \epsilon_- \equiv \Delta V$. Here, $a_\pm$ describes the different light degrees of freedom across the wall, and $T_\pm$ the different temperatures. One can explicitly compute the expression of the dof in the high--temperature limit, where they can be read from the thermal corrections to the effective potential:
\bea 
a_\pm={\pi^2 \over 30}\sum_{i=\text{light dof}} \l[ g^B_i+ {7 \over 8} g_i^F\r]\ , 
\eea 
where $B(F)$ stands for boson (fermion). From the EoS, it is easy to see that the relations in Eq.\eqref{eq:rel_velocities} become
\bea 
\label{eq:rel_velocities_bis}
v_+v_- = \frac{1-(1-3\alpha_+)r}{3-3(1+\alpha_+)r}\ ,\qquad \frac{v_+}{v_-} = \frac{3+(1-3\alpha_+)r}{1+3(1+\alpha_+)r}\ ,
\eea 
where we have defined 
\bea 
\label{eq:alpha_def}
\alpha_+ \equiv \frac{\epsilon_+ - \epsilon_-}{a_+ T_+^4}\ , \qquad r \equiv \frac{a_+T_+^4}{a_-T_-^4} \ ,  \qquad \alpha_N \equiv \frac{\epsilon_+ - \epsilon_-}{a_+ T_N^4}\ ,
\eea 
with $\alpha_N$ characterizing the strength of the PT at the nucleation temperature $T_N$. It is then conventional to define the vacuum energy in the true minimum to be zero: $\epsilon_- = 0$ and $\epsilon_+ \equiv \epsilon$. Notice that by doing so we are specifying our transition to proceed from a phase with a higher vacuum energy to a phase with a lower one. This is the usual behavior expected for a cooling phase transition, as it complies with the structure of the zero--temperature potential.

\begin{figure}
 \centering
 \includegraphics[width=0.48\textwidth]{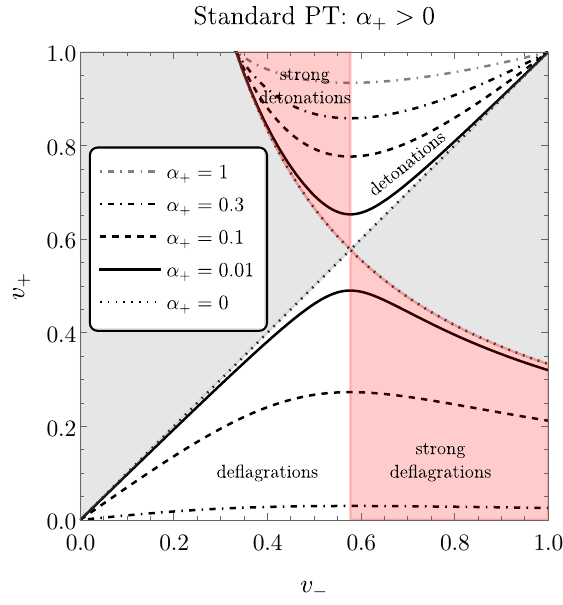}
  \includegraphics[width=0.48\textwidth]{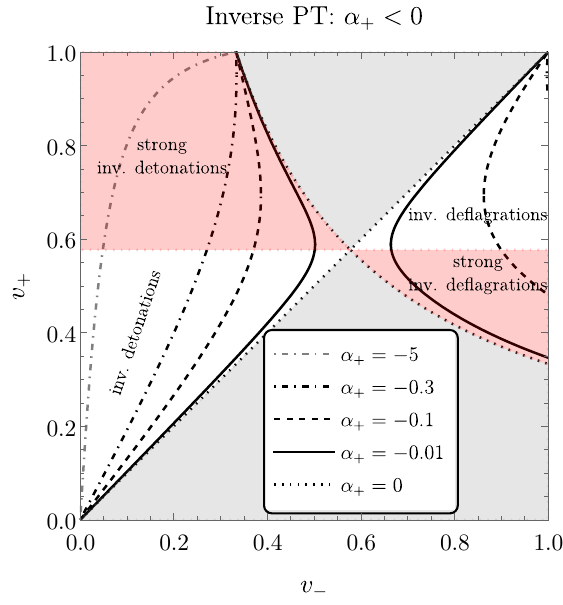}
 \caption{\textbf{Left}: In standard direct phase transitions with $\alpha_+>0$, we depict contours of constant $\alpha_+$ in the allowed region in the plane $(v_-, v_+)$, where $v_\pm$ are the fluid velocities in the wall frame. Shaded red regions indicate the presence of deflagrations and detonations, which are forbidden by hydrodynamical constraints, as we explain in the text. \textbf{Right}: Same as the left panel, but for the case of inverse phase transitions with $\alpha_+<0$. In the shaded red regions we similarly highlight the impossibility of strong inverse detonations ($v_+\leq c_s$), while strong inverse deflagrations will decay to inverse hybrids.}
 \label{fig:vpvm2}
 \end{figure}

The parameter $r$ can be eliminated from Eq.\eqref{eq:rel_velocities_bis} to write $v_+(v_-, \alpha_+)$, 
\bea 
\label{eq: v- in term of vp}
v_+(v_-, \alpha_+) = \frac{1}{1+\alpha_+} \bigg[\bigg(\frac{v_-}{2}+ \frac{1}{6v_-}\bigg) \pm \sqrt{\bigg(\frac{v_-}{2}+ \frac{1}{6v_-}\bigg)^2 + \alpha_+^2 +\frac{2}{3}\alpha_+- \frac{1}{3}}\bigg] \ .
\eea 
In Fig. \ref{fig:vpvm2} are reported the two different branches $\pm$ for constant values of $\alpha_+$. The left panel refers to direct PTs, with $\alpha_+>0$, while as we will discuss in Sec.\,\ref{sec:InvPT} the right panel refers to inverse PTs, with $\alpha_+<0$.

Phase transitions and discontinuities are accompanied by an increase in the entropy of the plasma. We discuss the conservation of entropy current 
\bea 
\partial_\mu (s u^\mu) = 0, \qquad s\equiv \frac{w}{T} \qquad \text{(Entropy in continuous waves)} \, .
\eea 
 This is a reflection of the fact that the fluid in a continuous wave is in local thermal equilibrium. However, discontinuous boundaries are intrinsically dissipative and entropy will generically increase across them. Across a boundary, the following inequality has to be fulfilled: 
\bea 
\label{eq:increase_of_entropy}
s_+ \gamma_+ v_+ \leq s_-\gamma_- v_-  \qquad \text{(Across discontinuities)} \, ,
\eea 
which imposes the increase of the entropy across the wall. 
Using the matching conditions in Eq.\eqref{eq:junctionAB}, this relation can also be rephrased in the following way 
\bea 
\frac{\gamma_-}{\gamma_+} \leq \frac{T_+}{T_-} \,. 
\eea

\subsection{The theory of discontinuities}
 So far we have addressed the hydrodynamic equations across discontinuities, focusing on matching conditions across the interfaces. The purpose here is to introduce the \emph{Taub} and \emph{reaction} adiabats that connect (and select) the physical fluid states across interfaces.
In the study of phase transitions, two types of discontinuities will be relevant: i) the shock front and ii) the reaction front (or phase boundary). Additional details on the construction of the adiabats are provided in Appendix \ref{app:taub_and_react_adiabat} as well as in Ref.\,\cite{RezBook}.

\paragraph{Shock front}
The first type of discontinuity that can develop in a fluid is the shock front. It is characterized by an interface where there is no change in the chemical/physical composition of the fluid (vacuum), but there can be discontinuous jumps in the thermodynamic quantities. 

\begin{figure}
    \centering
    \includegraphics[width=.48\textwidth]{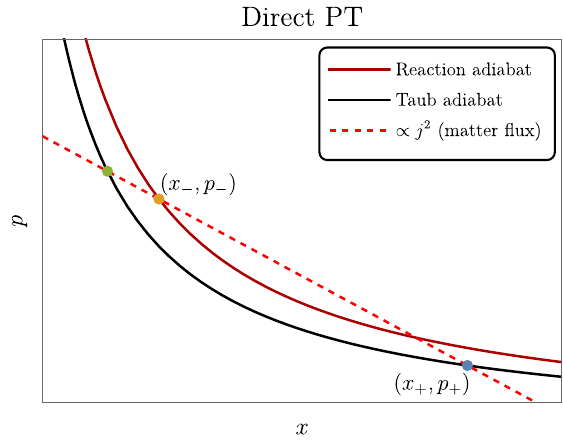}
    \includegraphics[width=.48\textwidth]{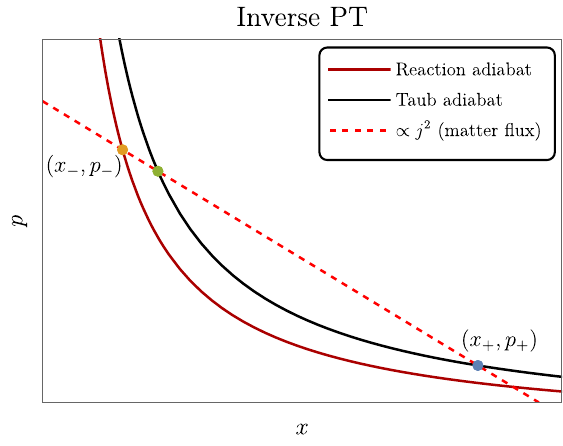}
    \caption{\textbf{Left}: Reaction and Taub adiabats for a direct PT. The trajectory connecting the blue dot with the green one is an example of a valid shock wave (connecting two points on the same Taub adiabat), while the trajectory connecting the blue dot to the orange one (from the Taub adiabat to the reaction one) is an example of valid reaction front with $\epsilon \neq 0$. Further details are presented in Appendix \ref{app:taub_and_react_adiabat}. \textbf{Right}: Same for the case of an inverse PT.}
    \label{fig:TaubVSRec}
\end{figure}

The matching conditions across a discontinuity, using Eq.\eqref{eq:conditionA} and Eq.\eqref{eq:conditionB}, can be expressed in the form 
\bea 
\label{eq:matching_Laine}
w_- x_- - w_+ x_+ = (p_- - p_+)(x_-+ x_+) \ ,
\eea 
where we defined $x \equiv w/\rho^2$, being $\rho$ the rest--mass density, which we find convenient\footnote{
In the case of nonrelativistic fluids, the shock adiabatic is typically plotted in the $(V,p)$-plane, with $V$ representing the volume of the system. However, for the relativistic case, the natural variables for representing the relativistic shock adiabatic are $w V^2=w/\rho^2$ and $pc^2$. Using these coordinates, the Taub adiabat offers a straightforward and graphical description of fluid properties across a shock. It can be visualized as the curve connecting the states ahead and behind a shock wave.} to represent in the plane $(x, p)$. Now we need to express the $w_-(x_-, p_-)$ as a function of $(x_-, p_-)$, where we consider $x$ as the independent variable, and this can be done via the bag equation of state, describing a system for which there is no change in its microscopic nature across the interface, that is $\epsilon=0$,
\bea 
w_- = 4 p_-\ , \qquad w_+ = 4 p_+ \ .
\eea 
Writing $p_-$ in terms of the other quantities we obtain
\bea 
p_- = \frac{3p_+x_+ -p_+ x_- }{(3x_- -x_+)} \ . \qquad \text{(Taub adiabat)}
\eea 
This is the \emph{Taub} adiabat. Intuitively, the shock front can be understood as a discontinuity with vanishing latent heat $\alpha_+ \to 0 \ (\epsilon\to 0)$, which implies, from Eq.\eqref{eq:rel_velocities_bis} 
\bea 
v_+ v_- = \frac{1}{3}\ , \qquad \frac{v_+}{v_-} = \frac{3+ \tilde r}{1 + 3 \tilde r}\ , \qquad \tilde r \equiv \bigg(\frac{T_+}{T_-}\bigg)^4\ ,
\eea 
since for a shock wave with no change of vev, the number of relativistic dof remains the same, i.e. $a_+ = a_-$. Finally, denoting the velocity of the shock wave $\xi_{\rm sh}$ and the velocity of the fluid after the shock wave $v_{\rm sh,-}$, we obtain the following matching conditions at the shock-wave front
\begin{align}
\xi_{\rm sh}=c_s\sqrt{\frac{3+\tilde{r}}{1+3\tilde{r}}}, \qquad
v_{\rm sh,-}=\frac{c_s^2}{\xi_{\rm sh}}\,.\label{eq:sw-matching2}
\end{align}
The state of the plasma across a shock wave has always to lie on the same Taub adiabat. Moreover, as described in Appendix \ref{app:taub_and_react_adiabat}, entropy considerations require that the state of the plasma behind the shock lies on a point of the Taub adiabat with larger pressure 
\bea 
\label{eq:shocks_entropy}
p_- > p_+ \qquad \text{(entropy increase for shocks)} \, , 
\eea 
or that the shock wave \emph{goes up on the Taub adiabat}, as for example illustrated on Fig.\ref{fig:TaubVSRec} for the trajectory from the blue to the green dot. 
\paragraph{Reaction front}
The second type of discontinuity is the reaction front, which is characterized by an interface where there is a change in the chemical/physical composition of the fluid. In other words, this requires a change in vacuum energy and in the equation of state, often manifested through a change in the number of relativistic degrees of freedom. 

Now, using the matching conditions across a discontinuity, Eq.\eqref{eq:conditionA} and Eq.\eqref{eq:conditionB}, and the bag equation of state, with $\epsilon\neq0$, we get
\bea 
\label{eq: reaction adiabat}
p_- = \frac{(3p_++4 \epsilon)x_+ -p_+ x_- }{(3x_- -x_+)}\ . \qquad \text{(reaction adiabat)}
\eea 
This is the \emph{reaction} adiabat. Let us notice that the reaction front clearly reduces to a shock wave in the limit $\epsilon \to 0$. Therefore, the reaction front connects states ahead of the interface, lying on the Taub adiabat, to states behind the interface, lying on the reaction adiabat.

In Fig.\ref{fig:TaubVSRec}, we present the Taub and the reaction adiabat with the solid black and darker red line, respectively. Starting from a point $(x_+, p_+)$ lying on the former adiabat, we can conclude that any other point on this adiabat can be reached upon crossing some shock wave. What selects the arrival point is the conserved flux going through the discontinuity. The matter flux $j$ is defined by 
\bea
\rho_+\gamma_+v_+=\rho_-\gamma_-v_-\equiv j \ ,
\eea 
where, in these coordinates, it provides the slope of the chord from the initial point + on the adiabatic to any other point --, that is:
\bea 
p_- = p_+ - j^2(x_- - x_+) \ ,
\eea 
and it is shown with the dashed red line in Fig.\ref{fig:TaubVSRec}. Given an initial point $(x_+, p_+)$ (the blue dot lying on the Taub adiabat) and a straight line with the (conserved) matter flux as the slope (the dashed red line), we can obtain the state of the plasma behind the wall as the intersection between such line and the reaction adiabat (the orange dot). This shows the intuitive usefulness of such construction.


\subsection{Hydrodynamical equations}

Up to this point, we have discussed the relations between thermodynamic quantities across discontinuities. In this section, our focus shifts to examining the different types of solutions to the hydrodynamical equations and investigating their properties.

The relativistic hydrodynamics equations have been shown to allow for \textit{self-similar} solutions. A hydrodynamic solution is said self-similar when it can be described by only two quantities with independent dimensions apart from space and time. In such cases, all relevant physical quantities can be expressed as functions of a similarity variable, typically a combination of spatial and temporal coordinates. 

For large enough bubbles, when the solution reaches a terminal wall velocity, the fluid profile can be characterized by the \emph{self-similar} variable $\xi \equiv r/t$ (for a comprehensive explanation, we refer to~\cite{Giulini2015LucianoRA, bookLandau, RezBook}). Notably, $\xi$ possesses the dimension of a velocity but can be interpreted as a position as well. The velocity of the bubble wall, denoted as $\xi_w$, ranges between the center of the bubble ($\xi \to 0$) and the lightcone ($\xi \to 1$). 

Starting from the conservation of the energy-momentum tensor, projecting Eq.\eqref{eq: energy cons} along and perpendicular to the flow, assuming spherical symmetry for the solutions and finally expressing the system of equations in terms of this self-similar variable, we obtain the following form 
\begin{align}
\label{eq:euler_conti}
(\xi - v ) \frac{\partial_\xi e}{w} &= 2 \frac{v}{\xi} + [1- \gamma^2 v (\xi - v) ] \partial_\xi v \ ,\qquad \text{(Euler eq.)}
\nn 
(1 - v \xi ) \frac{\partial_\xi p}{w} &= \gamma^2 (\xi -  v)\partial_\xi  v\ . \qquad \text{(continuity eq.)} 
\end{align}
We emphasize that $v$ denotes here the velocity of the fluid in the \emph{plasma frame} (equivalently the center of the bubble frame), as opposed to $v_{\pm}$ defined in the frame of the wall. 

Combining those equations leads to the well--known equation for the fluid velocity:
\bea 
\label{eq:fluidmotion}
2\frac{v}{\xi} = \gamma^2 (1-  v \xi) \bigg[\frac{\mu_{p \to w}^2}{c_s^2}-1\bigg] \partial_\xi v , \qquad \mu_{p \to w}(\xi,  v) = \frac{ v - \xi}{1-\xi v} \ ,
\eea
where $\mu_{p \to w}(\xi,  v)$ is the Lorentz transformed fluid velocity, from the plasma to the wall frame.
Two qualitatively different types of solutions emerge from the analysis of the equation Eq.\eqref{eq:fluidmotion}: i) the rarefaction wave and ii) the compression wave. 
\begin{enumerate}
    \item[i)] The rarefaction wave propagates from the head, moving at the largest velocity $\xi_{\rm head}$ to the tail, which moves at some smaller velocity $\xi_{\rm tail} < \xi_{\rm head}$. Consistency dictates that either $\xi_{\rm tail} = c_s$ or $\xi_{\rm head} = c_s$. While this is not true for generic flows, the symmetries of an expanding bubble dictate that the flow is at rest at the bubble center and on the lightcone, meaning $v(\xi =0) = v(\xi =1) = 0$. By imposing these boundary conditions, we deduce from Eq. \eqref{eq:fluidmotion} that for such solutions, $\partial_\xi v> 0$. Examining Eq. \eqref{eq:euler_conti} and focusing on the cases of interest for us, namely $\xi - v > 0$ and $\partial_\xi v>0$, we conclude that the rarefaction waves are also decompression waves, since $\partial_\xi p > 0$, as we travel from $\xi =1$ to $\xi =0$, i.e. toward the center of the bubble. Similar conclusions hold for the enthalpy $w$ and the temperature $T$, which we will define, in terms of the wall velocity, at the end of this section.
    \item[ii)] On the other hand, compression waves accelerate the motion of the plasma toward the center of the bubble. The same reasoning as above indicates that the pressure, the temperature, and the enthalpy increase across the wave, as we travel toward the center of the bubble, i.e. from $\xi= 1$ to $ \xi=0$.
\end{enumerate} 
Upon solving Eq. \eqref{eq:fluidmotion} with the matching condition \eqref{eq:conditionA}, and obtaining the fluid velocity profile, we can subsequently compute the enthalpy profile
\bea 
\label{eq: enthalpy}
w(\xi)=w(\xi_0)\,\exp \l[ \int_{v(\xi_0)}^{v(\xi)}\ \l( {1 \over c_s^2}+1\r) \gamma^2(v) \mu(\xi(v), v)\ dv \r] \ .
\eea 
From $\partial_\xi \ln T=\gamma^2(v) \mu(\xi, v) \partial_\xi v$, we can also obtain the temperature profile that reads
\bea 
\label{eq: temperature}
T(\xi)=T(\xi_0)\,\exp \l[ \int_{v(\xi_0)}^{v(\xi)}\ \gamma^2(v) \mu(\xi(v), v)\ dv \r] \ ,
\eea
where $\xi_0$ refers to the interface location, both for shock and reaction fronts. Properly computing the profile across the wall of the different thermodynamic quantities will be the subject of sec. \ref{sec:energy_budget}.

\subsection{The types of solutions for direct PTs}
\label{sec:types_of_solutions}
\begin{figure}
     \centering
     \includegraphics[width=.7\textwidth]{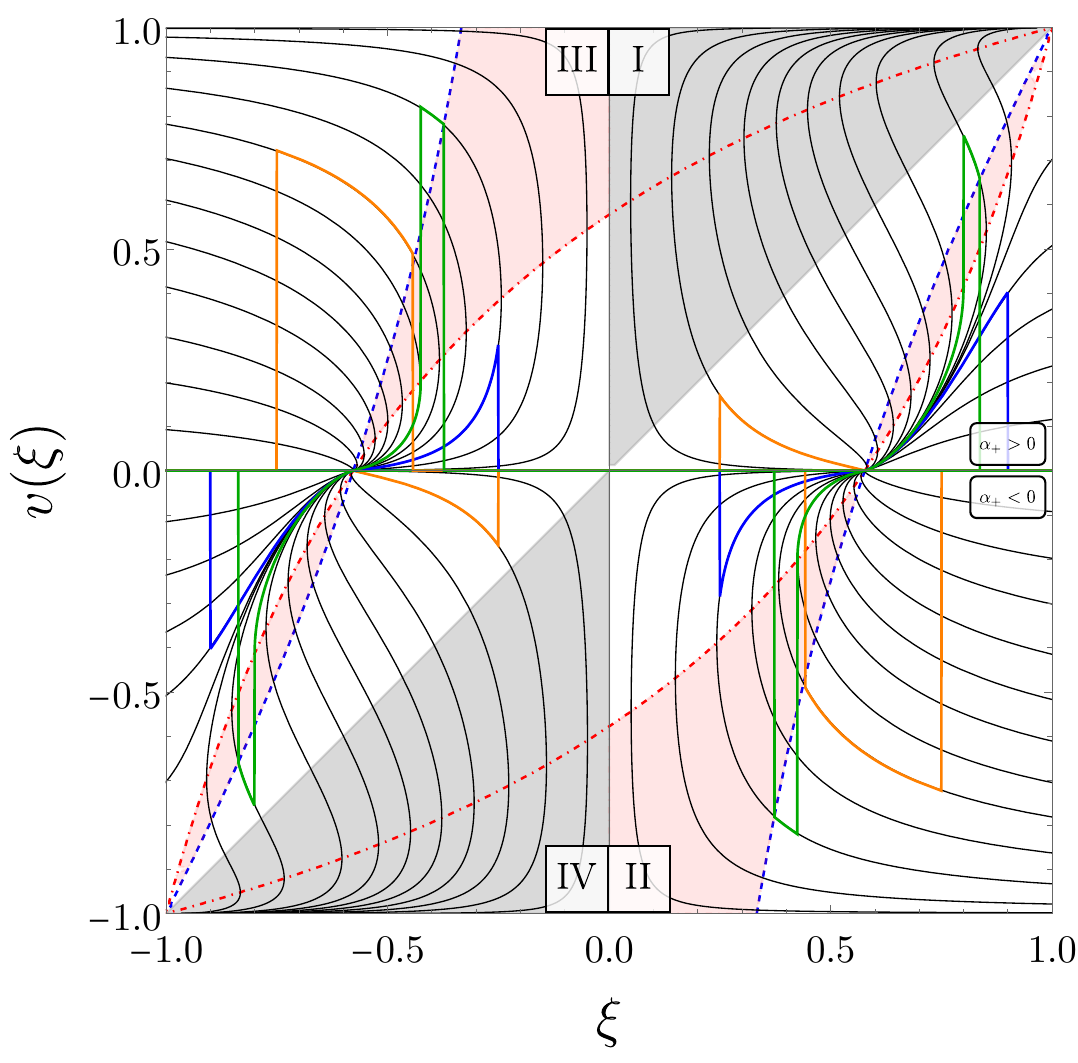}
     \caption{Profiles of the fluid velocity $v(\xi)$ in the plasma frame, both for the case of (standard) direct phase transitions ($\alpha_+>0$), and for inverse phase transitions ($\alpha_+<0$). The former case is described in the quadrant I and has $v(\xi)>0$, while the latter is in the quadrant II with $v(\xi)<0$. The gray shaded region is unphysical as it would imply that the fluid moves faster than the wall, $|v(\xi)|>|\xi|$. The red shaded region would similarly give unphysical velocity profiles as the dot--dashed red line indicates the maximum velocity that a detonation--type of solution can have, i.e. the sound speed in a frame moving at $\xi$, $v(\xi)=\mu(\xi, c_s)$. The dashed blue line shows the velocity of the shock front for deflagrations, that is $\mu(\xi_{sh}, v(\xi_{sh}))\xi_{sh}=c_s^2$. The different quadrants describe different physical systems: I) bubble and II) inverse bubble expansion as $\xi > 0$, III) droplet and IV) inverse droplet collapse with $\xi < 0$. The quadrants are related among each other by a mirror symmetry $v\to -v$ and $\xi \to -\xi$. The colored profiles in orange, green and blue, in the I quadrant, describe a deflagration, a hybrid, and a detonation, respectively, and in the II quadrant an inverse deflagration, an inverse-hybrid, and an inverse detonation, respectively. The other profiles are obtained by symmetry (see also\,\cite{RezBook}).}
     \label{fig:wave_def}
 \end{figure}

In the previous sections, we have gathered the necessary tools to solve the thermodynamic profiles of bubbles. We now apply these tools to various modes of expansion of cosmological bubbles.
As we shall see, constructing the velocity and temperature profiles of physical phase transitions requires gluing together discontinuous fronts and continuous waves.

In this section, we delineate the different types of solutions possible for direct cosmological phase transitions. In Fig. \ref{fig:wave_def} we present all the possible solutions of Eq. \eqref{eq:fluidmotion}. The four quadrants describe different physical situations: I) direct bubble expansion~\cite{Laine:1993ey, Laine:1994bf, Laine:1998jb}, II) inverse bubble expansion (this work), III) direct droplet collapse (see~\cite{Cutting:2022zgd} for a recent study) and IV) inverse droplet collapse. In this section we will remind the solutions in quadrant I and study in depth the solutions of quadrant II in Sec. \ref{sec:InvPT}. Table \ref{table_types_of_sol_direct} summarizes the various flows that can exist across the discontinuity. 

\begin{table}
\centering
\begin{tabular}{ |p{5cm}||p{5cm}||p{5cm}|  }
 \hline
 \multicolumn{3}{|c|}{Types of discontinuities for cosmological \emph{direct} phase transitions} \\
 \hline
& Detonations 
\newline
$p_+ < p_-, v_+ > v_-$& Deflagrations 
\newline
$p_+ > p_-, v_+ < v_-$\\
 \hline
Weak   & $v_+ > c_s, v_- > c_s$  Physical   & $v_+ < c_s, v_- < c_s$  Physical
 \\
 Chapman-Jouguet & $v_+ > c_s, v_- = c_s$  Physical &   $v_+ < c_s, v_- = c_s$  Physical \\
 Strong &    $v_+ > c_s, v_- < c_s$  Forbidden  & $v_+ < c_s, v_- > c_s$  Unstable\\
 \hline
\end{tabular}
\caption{Types of discontinuities for the direct phase transitions.}
\label{table_types_of_sol_direct}
\end{table}

Direct phase transitions admit three bubble expansion modes: i) detonations ii)  deflagrations and iii) hybrid solutions. 
\subsubsection{Detonations}

In phase transitions proceeding as detonations, the wall moves colliding with the fluid at rest in front of it. In the wall frame, the fluid ahead of the wall moves at $v_+=\xi_w$, and upon entering the new phase, it slows down to $v_- < v_+$. Detonations lie in in the upper half plane of the left panel of Fig.\,\ref{fig:vpvm2}.

To achieve a consistent solution, this setup needs a composition of a reaction front located at $\xi_w$, followed by a rarefaction wave. The fluid velocity right after the wall passage jumps to $v(\xi_w)=\mu(v_+, v_-)$ and then gradually slows down until it smoothly reaches zero at $\xi= c_s$. In this scenario, according to the junction conditions, we observe that $T_+ = T_{N} < T_-$ and $\alpha_+ = \alpha_N$. A family of solutions is parameterized by the parameter $r$, and we have indicated by $T_{N}$ the nucleation temperature at which the bubbles nucleate.

In principle, three possible classes of detonations exist depending on $v_-$. \emph{Weak} detonations exhibit $v_-> c_s$. In this case, the reaction front of the detonation is a \emph{weak discontinuity} and its trajectory on the reaction adiabat is depicted in the left panel of Fig.\ref{fig:whatW}. The \emph{Chapman-Jouguet} (CJ) detonations have $v_- = c_s$. On the other hand, \emph{strong} detonations have $v_- < c_s$, which display a strong discontinuity. The trajectory on the reaction adiabat of a strong discontinuity is presented in the left panel of Fig.\,\ref{fig:what}.

The velocity profile, in the plasma frame, for a weak detonation, is illustrated in the left panel of Fig.\,\ref{fig:profiles}. 
 
\paragraph{Impossibility of strong detonations}
In this paragraph, we review why strong detonations are not feasible for cosmological phase transitions. This impossibility stems from the boundary conditions, which impose that the velocity asymptotically approaches zero both far from the bubble and at its center~\cite{Laine:1993ey}. This requirement translates to $v(\xi = 0) =v(\xi \to c_s) \to 0$ and $v(\xi = 1)=0$. The resulting impossibility of strong detonation is readily apparent from Eq. \eqref{eq:fluidmotion}. Indeed, in the case of detonations, $v > 0$, and the decrease of velocity, $\partial_\xi v>0$, from the position of the front $\xi > c_s$ to $c_s$ is only possible if
\bea 
\bigg[\frac{\mu_{p \to w}^2}{c_s^2}-1\bigg]  > 0 \, ,
\eea 
which conversely requires that 
\bea 
|\mu_{p \to w}(\xi, v)| = \frac{\xi -v}{1-\xi v}  \equiv v_- > c_s \,. 
\eea
From these considerations, we conclude that strong detonations with $v_- < c_s$ cannot satisfy the boundary conditions of a bubble with vanishing bulk velocity at the center.

\subsubsection{Deflagrations}

For phase transitions described by deflagrations, the plasma is at rest immediately behind the wall, so the wall velocity is $\xi_w = v_-$. These solutions correspond to the lines in the lower half plane of the left panel in Fig.\,\ref{fig:vpvm2}, with the fluid velocity being higher behind the wall than in front, $v_- > v_+$. The fluid velocity in front of the wall jumps to $v(\xi_w) = \mu(v_-, v_+)$. Since $v_- > v_+$, we have $v(\xi_w) < \xi_w$, causing the deflagration solution profile to start below the line $v = \xi$, as shown in quadrant I of Fig.\,\ref{fig:wave_def}.

These solutions are constructed from the combination of a shock wave, followed by a compression wave, and finally, the reaction front, as depicted on the right panel of Fig.\,\ref{fig:profiles}. Similarly to detonations, deflagrations can, in principle, be weak $v_- < c_s$ (see right panel of Fig.\,\ref{fig:whatW} to see the trajectory on the reaction adiabats of the weak discontinuity), strong $v_- > c_s$ (right panel of Fig.\,\ref{fig:what}), or Chapman-Jouguet type $v_- = c_s$.

In the case of \emph{subsonic} deflagrations, the wall velocity is $v_- = \xi_w < c_s$, leading to the conclusion that these deflagrations are weak. However, this is not the case if the deflagration is supersonic (yet slower than the Jouguet velocity)~\cite{Kurki-Suonio:1995rrv}.
\paragraph{Impossibility of strong deflagrations} It has been argued that strong (supersonic) deflagrations are forbidden~\cite{Laine:1993ey,Kurki-Suonio:1995rrv}, primarily through two distinct arguments: stability and entropy law. The first argument, reviewed in Appendix \ref{app:stability}, concludes that even if a supersonic deflagration can indeed be made to exist at some moment, it is inherently unstable with respect to perturbations. The second argument relies on the increase of entropy. In Fig.\,\ref{fig:what}, we show how a \emph{strong} deflagration (left panel, blue to orange dot trajectory) can be seen as a shock (blue to green trajectory), followed by a \emph{weak} detonation (green to orange trajectory). It is instructive to compare with the case of a strong detonation (right panel, blue to orange trajectory), which can be seen as a shock (blue to green trajectory) followed by a weak deflagration (green to orange trajectory).

\emph{But what is the nature of the shock wave in each case?} For the strong detonation we observe that the shock (blue to green) goes up on the adiabat, constituting an \emph{entropy-increasing} shock wave and satisfying the conditions in Eq.\eqref{eq:shocks_entropy}. This makes the strong detonation seemingly viable\,\footnote{As mentioned before, strong detonations are however not compatible with the boundary conditions of the nucleated bubble.}. However, in the case of deflagrations, where we have a shock from blue to green, going down on the adiabat, it consequently decreases entropy and violates the conditions in Eq.\,\eqref{eq:shocks_entropy}. Due to these two reasons, we conclude that strong deflagrations are not viable modes of expansion of bubble walls. We however emphasize, as discussed in\,\cite{Kurki-Suonio:1995rrv}, that the nature of the cosmological phase transitions might allow for a direct jump between the blue and the orange point, making our argument inconclusive. In any case, we expect the stability argument to hold for cosmological phase transitions as well. 

\subsubsection{Hybrids}
\begin{figure}
 \centering
 \includegraphics[width=0.32\textwidth]{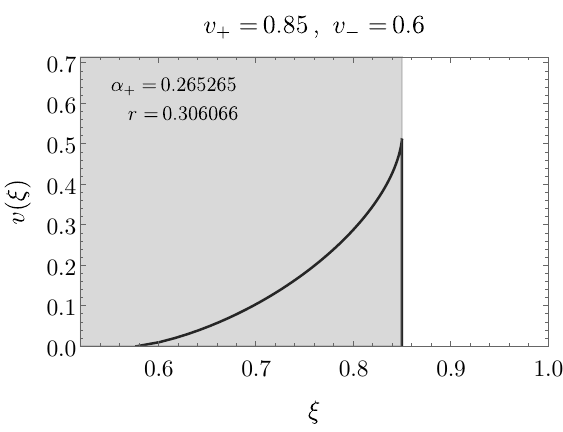}
  \includegraphics[width=0.315\textwidth]{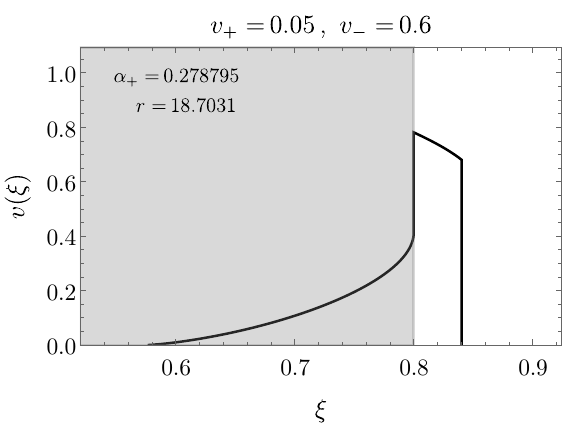} \includegraphics[width=0.32\textwidth]{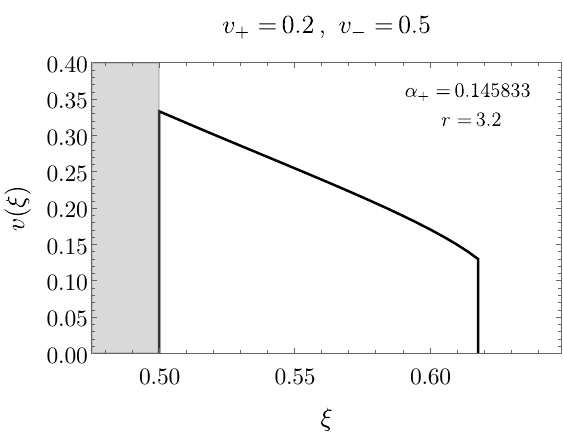}
 \caption{Velocity profiles for detonations (left), hybrids (middle) and deflagrations (right). The gray shaded region indicates the interior of the bubble. }
 \label{fig:profiles}
 \end{figure}

The previous reasoning by which strong deflagrations are forbidden does not necessarily mean that a supersonic deflagration--type of transition cannot exist. Although a solution with $v_- < c_s$ and $v_+ > c_s$ is unstable, the closest stable solution has $v_- = c_s$ and $v_+ > c_s$. This type of solution can be achieved by combining a Chapman-Jouguet deflagration with a rarefaction wave behind the wall, ensuring the fulfillment of boundary conditions. This configuration is termed \emph{hybrid solution} (or supersonic deflagration~\cite{Kurki-Suonio:1995rrv}), which become stable when $v_- = c_s$. In the middle panel of Fig.\ref{fig:profiles}, we present an example of such a hybrid composition.

\subsubsection{Evolving through the different profiles}

\begin{figure}
 \centering  \includegraphics[width=0.48\textwidth]{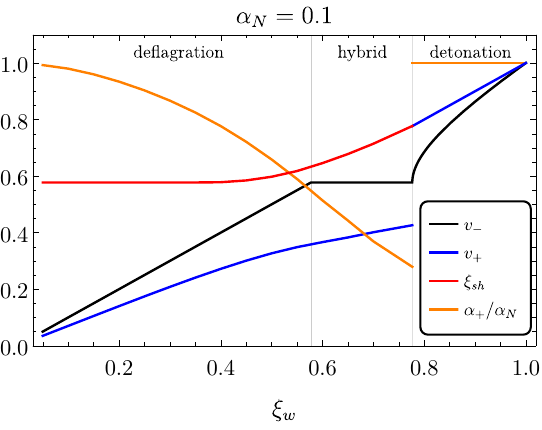}\ \includegraphics[width=0.488\textwidth]{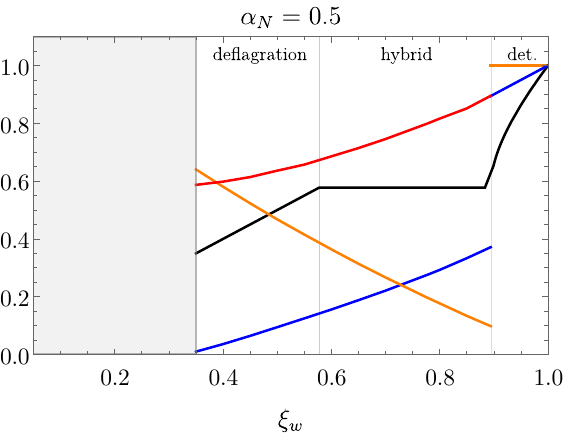}\\
 
 \caption{The quantities $\alpha_+/\alpha_N$, and $v_+, v_-, \xi_{sh}$ as a function of $\xi_w$ for fixed values of $\alpha_N = 0.1, 0.5$ varying the wall velocity. For deflagrations $\xi_w=v_-$, for hybrids neither $v_+$ nor $v_-$ can be identified as the wall velocity, while for detonations $\xi_w=v_+$. The vertical lines represent the sound speed and the Jouguet velocity. }
 \label{fig:quantities direct}
 \end{figure}

When the discussing the various modes of bubble expansion, it is more physical to fix the strength of the phase transition, $\alpha_N$, and solve for all the other quantities in terms of it. Having fixed the strength there will be only one possible fluid profile for a given wall velocity. In Fig.\,\ref{fig:quantities direct}, we present the evolution of $v_+$ ans $v_-$ as a function of $\xi_w$. Since $\alpha_N$ is not directly specified as an input parameter in the matching conditions, we conducted a scan over $v_+$ and $v_-$ to determine which combination gives the appropriate $\alpha_N$. Consequently, we also calculated $\alpha_+/\alpha_N$ and the position of the shock wave, $\xi_{sh}$. The vertical gray lines indicate the speed of sound, $c_s$, and the Jouguet velocity, which represents the velocity distinguishing the fastest hybrid solution from the slowest detonation. We will properly define the Jouguet velocity in section \ref{sec: jouguet}. In the right panel, we illustrate the scenario where for sufficiently large values of $\alpha_N$ a solution with $\xi_w$ within the gray band cannot be found. In other words, there cannot exist a (slow) deflagration compatible with such a (large) $\alpha_N$.
 
Now that we have examined the relevant characteristics of the solutions for an expanding bubble during a direct phase transition and we have set all the necessary notation, we are ready, in the next section, to explore the second quadrant of Fig.\,\ref{fig:wave_def}, which describes the case of bubble expansion during an inverse phase transition.

\section{Inverse phase transitions}
\label{sec:InvPT}

After reviewing the hydrodynamic solutions of direct PTs, we now turn to our main interest, namely the \emph{inverse} PTs. These occur from a phase with vanishing vacuum energy to another phase with (positive) vacuum energy.
We will follow the same presentation as before and identify all the expansion modes of inverse PTs.  

\subsection{Motivation and basics}


When the system finds itself in a minimum with zero vacuum energy, typically located away from the origin of the potential, it may still be advantageous to transition to a state with non--zero vacuum energy, albeit with more relativistic (light) degrees of freedom.

How can this be realized? The simplest setup is the one where a direct transition proceeds in the opposite direction, namely if we simply heat up the zero--temperature phase as considered in Ref.\,\cite{Buen-Abad:2023hex} in the context of reheating. If we stick to this picture, we do not need to go beyond the bag EoS. In fact, one can write the pressure in the broken and symmetric phases as in Eq.\,\eqref{eq:bag_eos} but 
inverting the roles of $\epsilon_{\pm}$, so that one can have a supercooled or superheated PT depending on which branch the system is coming from.
Notice that if the two phases are well--defined in the temperature range relevant for a direct transition, they are typically well--defined for the inverse transition as well, as both transitions are supposed to take place around the same critical temperature. Therefore, if the direct PT is not extremely supercooled, also the inverse PT should be well established provided it is not extremely superheated. 

The matching conditions of the inverse PT remain formally the same as in the direct case and we can refer to Sec.\,\ref{sec:matching_direct_PT}. As before, in order to complete the set of equations, we need to introduce an equation of state relating the various thermodynamic quantities. Keeping the same convention as in the previous section, for an inverse phase transition we require that the inside of the bubble $-$ expands into the outside of the bubble $+$, and the EoS takes the form in Eq.\,\eqref{eq:bag_eos}. Since the matching conditions across the wall are formally unchanged once they are written in terms of the enthalpy and the velocities in the wall frame, the relations in Eq.\,\eqref{eq:rel_velocities_bis} still hold. 

However, sticking to the definitions of Eq.\,\eqref{eq:alpha_def} for $\alpha_+$ and $r$, one finds that inverse PTs display negative $\alpha_+$:
\bea
\alpha_+ = \frac{\epsilon_+ - \epsilon_-}{a_+ T_+^4} = - \frac{\epsilon}{a_+ T_+^4}\ ,
\eea
where here $\epsilon_+ = 0$ and $\epsilon_- \equiv \epsilon$. From now on, we will stick to the following characterization: direct PTs have $\alpha_+ >0$ while inverse PTs have $\alpha_+ < 0$.

Following the same steps as before and eliminating the pressures and the $r$ parameter in favor of $\alpha_+$ and $v_-$, we obtain the same relation between the velocities 
\bea 
v_+(v_-, \alpha_+) = \frac{1}{1-|\alpha_+|} \bigg[\bigg(\frac{v_-}{2}+ \frac{1}{6v_-}\bigg) \pm \sqrt{\bigg(\frac{v_-}{2}+ \frac{1}{6v_-}\bigg)^2 + \alpha_+^2 -\frac{2}{3}|\alpha_+|- \frac{1}{3}}\bigg] \, , 
\eea 
with the only difference that $\alpha_+$ is now negative. Notice that the limit $\alpha_+ \to -1$ is actually smooth. 
The isocontours with constant $\alpha_+$ are reported in the right panel of Fig.\,\ref{fig:vpvm2}.

\subsection{The types of solutions for inverse PTs}
\begin{figure}
 \centering
 \includegraphics[width=0.32\textwidth]{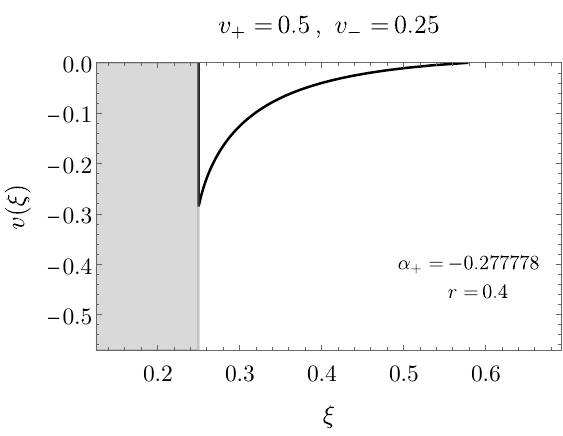} \includegraphics[width=0.32\textwidth]{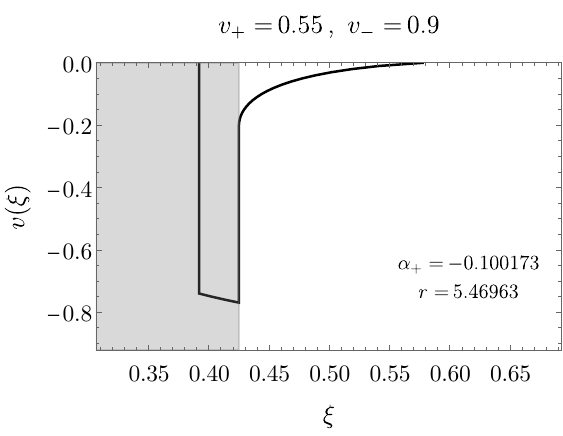} \includegraphics[width=0.32\textwidth]{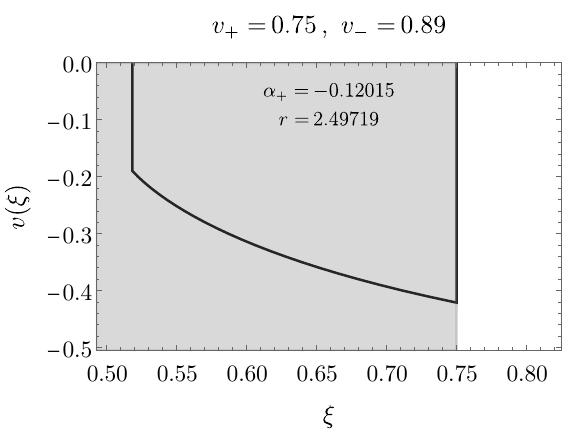}
 \caption{Velocity profiles for inverse detonations (left), inverse hybrids (middle) and inverse deflagrations (right).  }
 \label{fig:wave_def_bis}
 \end{figure}

\begin{table}
\centering
\begin{tabular}{ |p{5cm}||p{4cm}|p{4cm}|  }
 \hline
 \multicolumn{3}{|c|}{Types of discontinuities for cosmological \emph{inverse} phase transitions} \\
 \hline
& Inverse Detonations 
\newline
($p_+ < p_-, v_+ > v_-$)& Inverse Deflagrations 
\newline
($p_+ > p_-, v_+ < v_-$)\\
 \hline
Weak   & $v_+ < c_s, v_- < c_s$    & $v_+ > c_s, v_- > c_s$ 
 \\
 Chapman-Jouguet & $v_+ = c_s, v_- < c_s$ &   $v_+ = c_s, v_- > c_s$ \\
 Strong &    $v_+ > c_s, v_- < c_s$  & $v_+ < c_s, v_- > c_s$\\
 \hline
\end{tabular}
\label{table_types_of_sol_inv}
\caption{Types of discontinuity for the inverse phase transitions.}
\end{table}

Similarly to the case of direct phase transitions, we expect that several types of fluid solutions can exist for inverse phase transitions. We found five different possible expansion modes, analogously to the direct case, that we called: i) inverse detonations (weak and CJ), ii) inverse deflagrations (weak and CJ), and iii) inverse hybrids, displayed in the left, right, and middle panels, respectively, of Fig.\,\ref{fig:wave_def_bis}. 
Our naming of \emph{inverse} detonations and deflagrations relies on the mirror symmetry that can be drawn from Fig.\,\ref{fig:wave_def}
\footnote{The distinguishing physical characteristic of detonations setting it apart from deflagrations, as stated in \cite{Heinz1938}, is that the fluid just behind the reaction front is in motion rather than the propagation exceeding the speed of sound. The mirror symmetry flips this physical interpretation, as for instance for inverse detonations the fluid will be in motion ahead of the reaction front.}.

\subsubsection{Inverse Detonations}
The first possibility, in analogy with detonations, would be to build an \emph{inverse detonation} by gluing a reaction front with $\xi_w = v_- < c_s$ and a rarefaction wave going from $v(\xi_w^+)$ to $0$ at $\xi = c_s$. In the plasma frame, the velocity $v(\xi_w^+)=\mu(v_-, v_+)$ with $v_-<v_+$ is always negative. Notice however that since $\xi$ is positive the bubble is actually expanding. As one can see from Eq.\,\eqref{eq:euler_conti}, across the rarefaction wave, namely from $\xi=c_s$ to $\xi = \xi_w$, the pressure as well as the velocity decreases. Such solutions are displayed in the left panel of Fig.\,\ref{fig:wave_def_bis}.

For this type of wave with $v_- = \xi_w < c_s$, there are in principle three solutions for $v_+$, see the right panel of Fig.\,\ref{fig:vpvm2}: $v_+ < c_s$, which we call the \emph{weak} inverse detonation, the CJ solution $v_+ = c_s$ and $v_+ > c_s$, which we call the \emph{strong} inverse detonation.
The corresponding trajectories in terms of reaction adiabats are shown in Fig.\,\ref{fig:whatA} and  Fig.\ref{fig:whatWA}.


\paragraph{Impossibility of strong inverse detonations} Similarly to the case of direct strong detonations, one can show that inverse strong detonations are not compatible with the boundary conditions of the bubble. Taking into account that the fluid velocity is always negative, $v(\xi) \leq 0$, and requiring that $ v(\xi \to c_s) \to 0$, one has $\partial_\xi v > 0$. From Eq.~\eqref{eq:fluidmotion} this implies that the velocity increases from the phase boundary to the outside of the bubble, so that  $\l[\frac{\mu^2_{p \to w}}{c_s^2}-1\r] <0 $ is required by the boundary conditions. This yields
\bea 
|\mu_{p \to w}| = \frac{\xi -v}{1-\xi v}  \equiv v_+ < c_s \, ,
\eea
where the subscript $p \to w$ means that we go from the plasma frame to the wall frame. 
We then conclude that strong inverse detonations are not compatible with our boundary conditions. This leaves the weak and CJ inverse detonations as the only physical solutions. 

\subsubsection{Inverse Deflagrations}
A second type of solutions are \emph{inverse deflagrations}, where we glue a reaction front with $v_+ = \xi_w$ and $v_- > c_s$ followed by a compression wave ending with a shock front.
A typical fluid profile can be seen in the right panel of Fig.\,\ref{fig:wave_def_bis}. 
As we can see, these solutions are characterized by the fluid being sucked inside the bubble.
This can be physically interpreted by noticing that, upon entering the bubble, particles losing their mass receive a kick toward the center (as we shall see in Sec.\,\ref{sec:friction_inv_PT}), which then accelerates the fluid inward.

There are three possible solutions for inverse deflagrations depending on the value of $v_+$: $v_+ > c_s$ dubbed \emph{weak} inverse deflagration, the CJ solution with $v_+ = c_s$, and the \emph{strong} inverse deflagration with $v_+ < c_s$.
The corresponding trajectories in terms of reaction adiabats can be found in Fig.\,\ref{fig:whatA}
and Fig.\,\ref{fig:whatWA}.

\paragraph{The impossibility of strong inverse deflagrations} As for the case of direct PTs, strong inverse deflagrations can be argued to be forbidden by two arguments. 

The first argument is about stability, and
in Appendix \ref{app:stability} we argue that strong inverse deflagrations cannot be stable by following the same reasoning as for direct PTs. 

As a second argument, we can consider the behavior of entropy across a strong inverse deflagration front. Referring to the right panel of Fig.\,\ref{fig:whatWA}, we can interpret a strong inverse deflagration (blue to orange) as composed by a shock wave (blue to green) followed by a weak deflagration (green to orange). This is however forbidden, as the shock would imply negative entropy production (the pressure decreases across the shock)\footnote{Similar considerations would show that strong inverse detonations are instead allowed in principle, even though incompatible with the required boundary conditions for the bubble as discussed above.}. We again warn the reader that this second argument might not hold because of the nature of cosmological PTs.

\subsubsection{Inverse Hybrids} \label{sec: inv hyb} The closer stable solution to a strong inverse deflagration is a CJ inverse deflagration with $v_+ = c_s$, which we dub \emph{inverse hybrid}. Notice that here, contrarily to direct hybrids, it is the upstream velocity ($v_+$) that is fixed to the speed of sound. To build the velocity profile we glue a rarefaction wave to a detonation front, followed by a compression wave and finally a shock wave. This is shown in the middle panel of Fig.\,\ref{fig:wave_def_bis}. Moreover, the requirement that the shock wave is evolutionary (increases entropy) with $v_{-, sh} < c_s$ and $v_{+, sh} > c_s$, in the shock frame, imposes that $v_- > c_s$\footnote{Notice that in principle, another hybrid solution exists with $v_- < c_s$ and $v_+ = c_s$. However, this solution would require a shock wave behind the wall with $v_{sh,-}< c_s$ and $v_{sh,+}< c_s$, which is however forbidden by the thermodynamics of shocks. }. 

Fixing $v_+ = c_s$ actually sets an upper bound on $|\alpha_+|$. In particular, we find that for $|\alpha_+|\geq2/\sqrt{3}-1\sim 0.15$ inverse hybrids cannot be achieved. 
This is because these solutions are constrained to be in the region dubbed ``strong inverse deflagration" in Fig.\,\ref{fig:vpvm2},
and by writing $\alpha_+$ in terms of $v_+$ and $v_-$, we get from \eqref{eq: v- in term of vp} that
\bea 
\label{eq:alpha_v_+_v_rel}
\alpha_+(v_+, v_-)= {(v_--v_+)(1-3v_+v_-) \over 3v_-(1-v_+^2)} \ .
\eea 
By setting $v_+ = c_s$, and the extremal value $v_- = 1$ in Eq.\,\eqref{eq:alpha_v_+_v_rel}, we obtain the largest possible $|\alpha_+|$ as $|\alpha^{\text{max, hybrid}}_+| = |\alpha_+(c_s, 1)|$.

We also find that the wall velocity of the inverse hybrid solution must satisfy 
\bea 
c_s^2< \xi_w <c_s \ .
\eea 
 The lower bound comes from the fact that the slowest possible inverse hybrid is set by the slowest possible shock, that is the intersection of the blue dashed line with the horizontal axis $v(\xi)=-1$ in the quadrant II of Fig.\,\ref{fig:wave_def}. This reads
\bea 
\label{eq:slowest_hybrid}
\mu(\xi_w, v(\xi_w))\xi_w=c_s^2 \quad \overset{v=-1}{\longrightarrow} \quad \xi^{\text{min, hybrid}}_w=c_s^2\ .
\eea 
The upper bound comes from the fact that for $\xi_w >c_s$ there always exists an inverse deflagration which is stable.

\begin{figure}
 \centering 
 \includegraphics[width=0.48\textwidth]{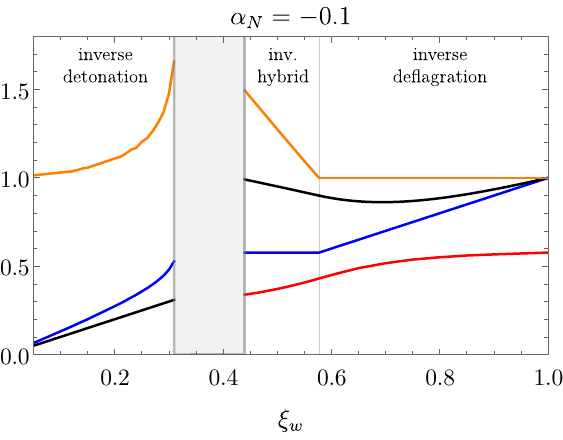}
 \includegraphics[width=0.48\textwidth]{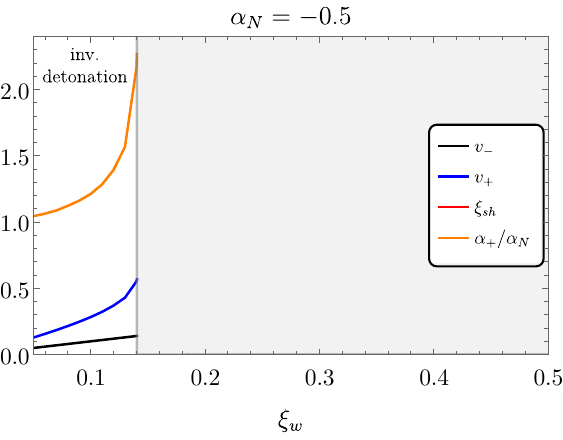}\\
 
 \caption{The quantities $\alpha_+/\alpha_N$, and $v_+, v_-, \xi_{sh}$ for fixed $\alpha_N = -0.1, -0.5$ varying the wall velocity. The color coding is the same as in Fig. \ref{fig:quantities direct}. For inverse detonations we can identify $\xi_w=v_-$, for inverse hybrids neither $v_+$ nor $v_-$ can be identified with the wall speed, while for inverse deflagrations $\xi_w=v_+$. We can see that for large enough $|\alpha_N|$, a window opens up where no consistent solution can be found. For $\alpha_N<-1/3$, as in the right panel, only inverse detonations are allowed.}
 \label{fig:quantities}
 \end{figure}

\subsection{The Jouguet velocity for the transition between detonation and hybrid}
\label{sec: jouguet}

In the context of direct phase transitions, the Jouguet velocity defines the wall velocity, $\xi_w= \xi^{\rm direct}_J $, for which the hybrid solution becomes a detonation, i.e. the velocity for which the bubble wall catches up with the shock wave. It reads 
\begin{align}
\label{eq:xiw-hyb}
     \xi^{\rm direct}_J=c_s\left(\frac{1+\sqrt{3\alpha_N\left(1-c_s^2+3c_s^2 \alpha_N\right)}}{1+3 c_s^2\alpha_N}\right),
\end{align}
which is obtained by substituting $v_-=c_s$ into Eq.\eqref{eq: v- in term of vp}, and we have identified $\alpha_+\equiv \alpha_N$ since the fluid to the right of the wall is at rest.  

By a similar reasoning, in this section we determine the inverse Jouguet velocity, $\xi_J^{\rm inv}$, which separates a pure inverse detonation from an inverse hybrid solution so that if $\xi_J^{\rm inv} < \xi_w < c_s$ the bubble grows as an inverse hybrid. 
To this end, we remind the reader that for an inverse detonation one has 
\bea 
v_- = \xi_w < v_+< c_s,
\eea 
while the inverse hybrid solution implies 
\bea 
v_- > c_s\ , \qquad v_+ = c_s\ , \qquad c_s^2< \xi_w <c_s \,. 
\eea 
We can now look for the wall velocity such that inverse detonations become impossible.  Starting from $v_+ < c_s$ for a weak inverse detonation, the transition to a \emph{forbidden} strong inverse detonation occurs when $v_+ > c_s$, so that the fastest allowed inverse detonation is given by the limit $v_+ \to c_s$. To find $\xi^{\rm inv}_J$ we can then set $v_- = \xi^{\rm inv}_J$ and $v_+ = c_s$ in Eq.\,\eqref{eq: v- in term of vp} to obtain
\bea 
\label{eq: v- in term of vp_bis}
c_s = \frac{1}{1-|\alpha_+|} \bigg[\bigg(\frac{\xi^{\rm inv}_J}{2}+ \frac{1}{6\xi^{\rm inv}_J}\bigg) -\sqrt{\bigg(\frac{\xi^{\rm inv}_J}{2}+ \frac{1}{6\xi^{\rm inv}_J}\bigg)^2 + \alpha_+^2 -\frac{2}{3}|\alpha_+|- \frac{1}{3}}\bigg],
\eea 
which implicitly gives $\xi_J^{\rm inv}$ as a function of $|\alpha_+|$. 

Notice that for inverse hybrid solution the region in front of the wall has a non--vanishing fluid velocity, and one cannot identify $\alpha_+$ with $\alpha_N$. The latter needs to be solved for once we have fixed the strength of the PT. The value of $\xi_J^{\rm inv}$ as a function of the wall velocity resulting from this procedure is shown by the red dashed line in the right panel of Fig.\,\ref{fig:efficiency}.

In Fig.\,\ref{fig:quantities}, we show the evolution of $v_+, v_-$ (defined at the reaction front), the ratio $\alpha_+/\alpha_N$, and the position of the shock $\xi_{sh}$, as a function of the wall velocity $\xi_w$ for the case of inverse phase transitions. Here we observe a gap of velocities between inverse detonations and inverse hybrids that cannot be realized for a given value of $\alpha_N$. The gap widens as $|\alpha_N|$ increases, and for $\alpha_N<-1/3$ the forbidden region extends to all the inverse hybrids and inverse deflagrations, leaving the inverse detonations as the only possible solutions.

\section{Energy budget of phase transitions}
\label{sec:energy_budget}

In this section, we examine the energy budget of inverse PTs. We begin by reviewing the direct PT case, and calculating the enthalpy and other thermodynamic quantities for the various solutions. We then compute the efficiency factor, which quantifies how much of the available energy is converted into bulk fluid motion, thereby contributing to the production of gravitational waves (GWs). Finally, we will present a similar analysis for inverse PTs and highlight the main differences.

\subsection{Thermodynamic quantities for direct PTs}
\begin{figure}
    \centering
    \includegraphics[width=.33\textwidth]{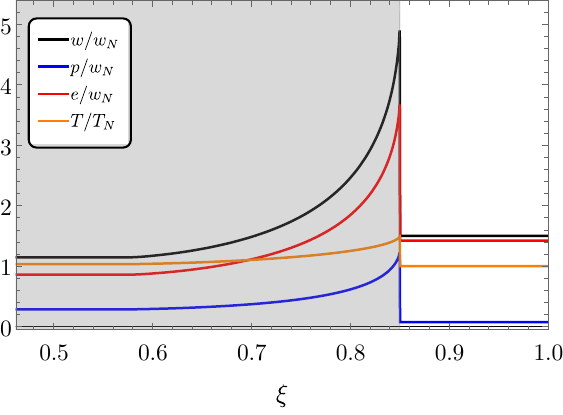}\includegraphics[width=.33\textwidth]{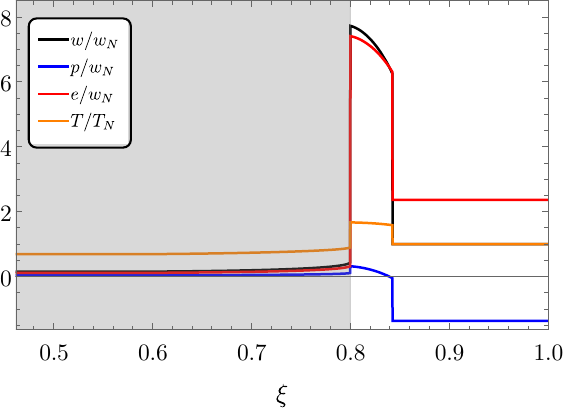}\includegraphics[width=.33\textwidth]{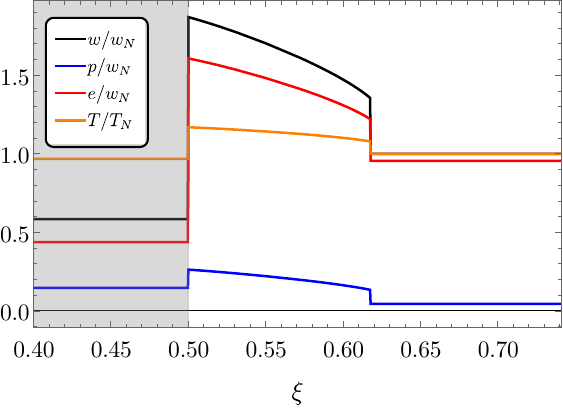}\\\includegraphics[width=.33\textwidth]{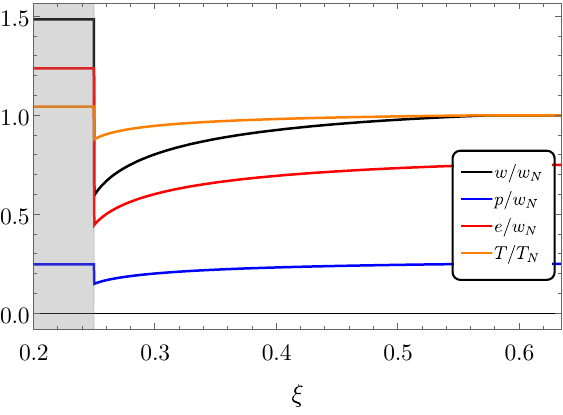}\includegraphics[width=.33\textwidth]{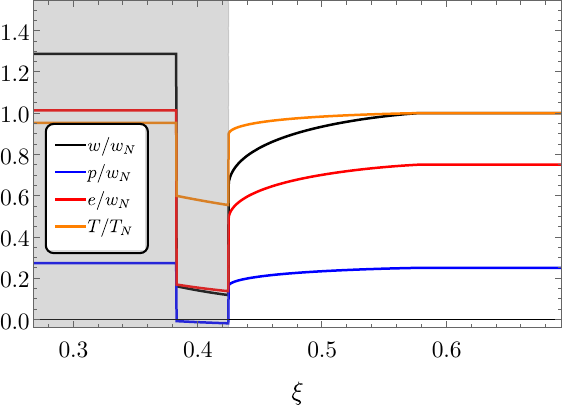}\includegraphics[width=.33\textwidth]{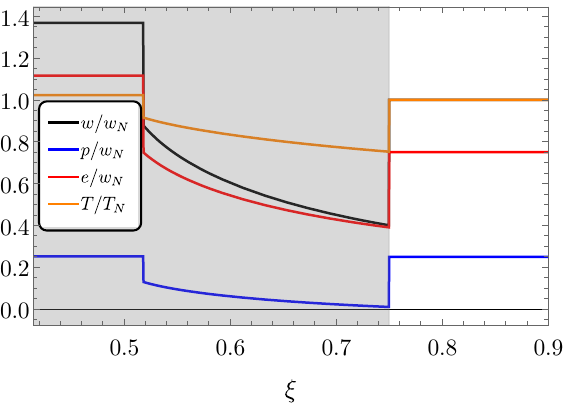}
    \caption{\textbf{Top}: Enthalpy, pressure, energy and temperature profiles across the bubble wall for detonations (left), hybrids (middle), and deflagrations (right). These correspond to the velocity profiles shown in Fig.\,\ref{fig:profiles}. The enthalpy and the temperature are normalised to $w_N$ and $T_N$. For the temperature profiles, we have chosen $a_+/a_-=1.5$. \textbf{Bottom}: Same as in the top row but for inverse PTs, for solutions corresponding to the profiles in Fig.\,\ref{fig:wave_def_bis}. For the temperature profiles, we have chosen $a_+/a_-=0.8$.}
    \label{fig:enthalpy profiles}
\end{figure}

As discussed already in Sec.\,\ref{sec:direct_PT}, once the fluid velocity profile is known it is possible to build the corresponding enthalpy and the temperature profiles. As these profiles are discontinuous at the wall and at the position of the shock, we need to use the junction condition in \eqref{eq:conditionA} to provide a full characterization. We consider the three types of solutions for the fluid and collect all the necessary results in Appendix \ref{app:thermo}. The resulting profiles are shown in the top row of Fig.\,\ref{fig:enthalpy profiles}.

\paragraph{Efficiency factor}

 The conservation of the energy momentum tensor from Eq.\,\eqref{eq: energy cons} tells us how much of the initial energy is converted into bulk fluid motion. We are interested in such a quantity because the kinetic energy density of the fluid, given by 
 \bea
\rho_{\rm kin} \propto v^2 \gamma^2 w \, ,
\eea
 controls the amplitude of the GW signal from the PT. It is useful to split the conservation of energy in the following way:
\bea
\label{eq: energy budget standard}
\underbrace{\frac{\xi_w^3}{3} \epsilon}_{\text{vacuum energy}} +\underbrace{\frac{3}{4} \int w_N \xi^2 d \xi}_{\text{initial thermal energy}} = \underbrace{\int \gamma^2 v^2 w \xi^2 d \xi}_{\text{fluid motion}}+ \underbrace{\frac{3}{4} \int w \xi^2 d \xi}_{\text{final thermal energy}} \ ,
\eea
The integration range in Eq.~\eqref{eq: energy budget standard} needs to include all the region of space where the fluid is perturbed ($v \neq 0$) for the conservation to hold.
This is interpreted as the fact that the released vacuum energy and the initial thermal energy are converted into bulk fluid kinetic energy and thermal energy after nucleation. By defining
\bea
\rho_N \equiv \frac{3}{4} \int w_N \xi^2 d \xi   \, ,\qquad \rho_{\rm kin} = \int \gamma^2 v^2 w \xi^2 d \xi\, ,
\eea
one can introduce the \emph{efficiency} parameter $\kappa_{\rm direct}$, 
\begin{equation}
    {\rho_{\rm kin}\over \rho_{\rm tot}}\equiv \kappa_{\rm direct} {  \alpha_N  \over 1+ \alpha_N}\ ,
\end{equation}
where $\rho_{\rm tot}$ is the total energy before nucleation, i.e. the LHS in Eq.\,\eqref{eq: energy budget standard}, and we have related the various quantities to $\alpha_N$ defined in Eq.\,\eqref{eq:alpha_def} via the bag EoS. The parameter $\kappa_{\rm direct}$ defined in this way is a measure of the efficiency for transferring the initial vacuum energy into bulk fluid motion and takes the form
\bea 
\label{eq:exp_kappa}
\kappa_{\rm direct}= {3\over \epsilon \xi_w^3} \int \gamma^2 v^2 w \xi^2 d \xi \ .
\eea 
The fraction of the total energy going into thermal energy can then be directly estimated by $1- \kappa_{\rm direct}$. 

\begin{figure}
    \centering
    \includegraphics[width=.48\textwidth]{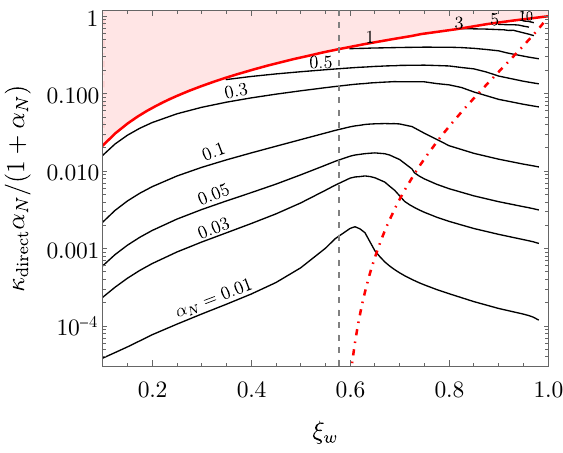}\includegraphics[width=.48\textwidth]{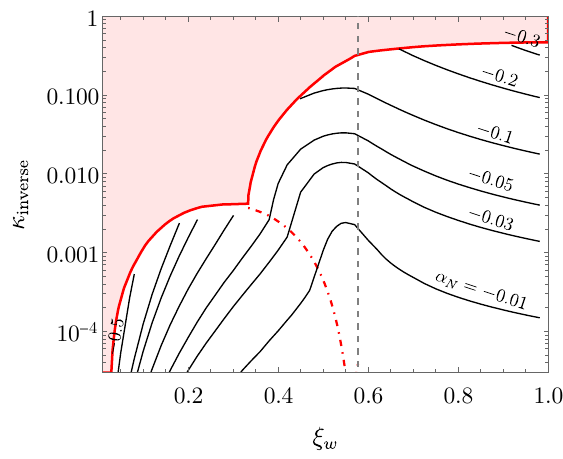}
    \caption{Efficiency factor for converting the energy budget into bulk fluid motion for direct PTs (left) and in the inverse case (right). The red dashed line is the Jouguet velocity. The red shaded region shows where no consistent solution with such a choice of wall velocity and efficiency can be found.}
    \label{fig:efficiency}
\end{figure}

The numerical results for $\kappa_{\rm direct}$ are displayed in the left panel of Fig.\,\ref{fig:efficiency}, where we actually plot the combination $\kappa_{\rm direct} \,\alpha_N/(1+\alpha_N)$ for a better comparison with the inverse case. The solid black lines are the isocontours with the same $\alpha_N$, varying the wall velocity. We can note that the top right corner, where we have the fastest hybrid solutions, is the most efficient in converting the released vacuum energy into kinetic energy, saturating almost to $1$ for $\alpha_N\gg1$. The red dot-dashed line is the Jouguet velocity, defined in Eq. \eqref{eq:xiw-hyb}, while the gray dashed line indicates the sound speed. It is worth noticing that not for every set of parameters a solution with a specific efficiency can be found. This is represented by the red shaded region, where no solution is available. This behavior is similar to the right panel of Fig.\,\ref{fig:quantities direct}, where the gray shaded region was signaling that no solution could be found for a certain choice of $\alpha_N$ and $\xi_w$. 

We also checked numerically energy conservation, Eq.\,\eqref{eq: energy budget standard}, which can be rewritten as
\bea 
\int_0^1 \l[ \l(\gamma^2-{1 \over 4}\r) w- {3\over 4} w_N\r]\xi^2 \ d\xi= {\epsilon\over 3}\xi_w^3 \ .
\eea

\subsection{Thermodynamic quantities for inverse PTs}
In this section, we aim to build the profiles for the thermodynamic quantities of interest for the inverse PTs. All the details are collected in Appendix \ref{app:thermo} and the profiles for the inverse transitions are presented in the bottom row of Fig.\,\ref{fig:enthalpy profiles}.

\paragraph{Efficiency factor}For the inverse PTs, we start by considering the energy density before the nucleation event to be the one of radiation,
\bea
\rho_{\rm tot} = e_R = \frac{3}{4} w_N \, ,
\eea
where $w_N$ is the enthalpy of the + phase at the nucleation temperature.
On the other hand, the kinetic energy density of the fluid is given by
\bea
\rho_{\rm kin} \propto v^2 \gamma^2 w \, .
\eea
It is again instructive to split the conservation of energy. For the inverse PTs, we obtain
\bea
\label{eq: energy budget inverse}
\underbrace{\frac{3}{4} \int w_N \xi^2 d \xi}_{\text{initial thermal energy}} = \underbrace{\frac{\xi_w^3}{3} \epsilon}_{\text{vacuum energy}} +\underbrace{\int \gamma^2 v^2 w \xi^2 d \xi}_{\text{fluid motion}}+ \underbrace{\frac{3}{4} \int w \xi^2 d \xi}_{\text{final thermal energy}} \ ,
\eea
where wrt to the standard case we see that the total amount of energy at our disposal is the initial thermal energy that will be converted not only into kinetic and final thermal energy, but also into vacuum energy. Indeed, it is apparent from Eq.~\eqref{eq: energy budget inverse} that inverse phase transitions are happening ``against the vacuum" and would not be possible at zero temperature.

In order to understand what are the appropriate boundaries of integration, we can consider the total enthalpy before nucleation inside a sphere that will contain all the space affected by the fluid perturbation after nucleation. This sphere has radius $\bar v = \,\,\text{Max}(\xi_w, c_s)$, where for instance $\bar v = c_s$ for inverse detonations.

Then, it is possible to define the efficiency $\kappa_{\rm inverse}$ as the fraction of the critical energy inside this sphere that will go into bulk fluid motion:
\bea
 \kappa_{\rm inverse} \equiv {\rho_{\rm kin}\over \rho_{\rm tot}}= \frac{4\pi \int_0^{\bar v} \xi^2 d \xi \,\,v^2 \gamma^2 w}{4\pi \int_0^{\bar v} \xi^2 d \xi \, \frac{3}{4} w_N} = \frac{4}{\bar v^3} \int \xi^2 d\xi \, v^2 \gamma^2 \frac{w}{w_N}\ .
\eea
In the way it is defined the efficiency $\kappa_{\rm inverse}$ is the parameter directly entering the fits for the GW signal\,\footnote{We could in principle also introduce another definition of the efficiency, $\tilde \kappa_{\rm inverse}$, which matches more closely the expression for the direct case:
\bea
\tilde \kappa_{\rm inverse} = \frac{3}{\epsilon \, \bar v^3} \int \xi^2 d\xi \, w \,v^2 \gamma^2\, , \quad
\kappa_{\rm inverse} = |\alpha_N| \tilde \kappa_{\rm inverse}.
\eea
}.
As before, we fix $\alpha_N$ and solve for all the other variables. Our numerical results for $\kappa_{\rm inverse}$ obtained in this way as a function of $\xi_w$ and $\alpha_N$ are displayed in the right panel of Fig.\,\ref{fig:efficiency}. The solid black lines are the isocontours with the same $\alpha_N$, varying the wall velocity. 

Similarly to the direct PT case, we note that the top right corner where we have the fastest inverse deflagrations is the most efficient in converting the initial energy into kinetic energy, saturating to $\kappa_{\rm inverse} \approx 0.5$ for $\alpha_N \gtrsim -1/3$. For more negative values of $\alpha_N$, i.e. $\alpha_N \lesssim -1/3$, only inverse detonations are allowed and the efficiency drops. This effect may also be understood in terms of energy conservation, as for large and negative $\alpha_N$ a significant fraction of the energy budget is lost to the vacuum energy of the new phase and is not transferred into bulk kinetic energy of the fluid.

The red dot-dashed line is the Jouguet velocity, defined implicitly in Eq.\,\eqref{eq: v- in term of vp_bis}, while the gray dashed line indicates $\xi_w = c_s$. The limiting solid red curve separates physical solutions from forbidden ones, in shaded red. It is worth noticing that the region for which a consistent solution cannot be found is bigger than in the direct case. Indeed, increasing $|\alpha_N|$ beyond $|\alpha_N| \approx 0.07$, a window of forbidden solutions appears between the inverse detonation and inverse hybrid, as we can already observed on Fig.\ref{fig:quantities}. This window induces the peculiar feature in the limiting red curve appearing at $|\alpha_N|\gtrsim 0.07$. At the dip, in $|\alpha_N|\gtrsim 0.07$ and $\xi_w = c_s^2$, it merges with the Jouguet velocity (dashed red line). This merging is due to the fact that the slowest possible velocity for an hybrid is $\xi_w=c_s^2$, 
as explained in section \ref{sec: inv hyb}.

We have also checked numerically energy conservation from \eqref{eq: energy budget inverse} for the inverse PT:
\bea 
\int_0^1 \l[ \l(\gamma^2-{1 \over 4}\r) w- {3\over 4} w_N\r]\xi^2 \ d\xi=- {\epsilon\over 3}\xi_w^3 \ .
\eea

\subsection{The GW signal from sound waves}

One of the main interesting features of cosmological phase transition is the copious gravitational wave signal induced~\cite{Witten:1984rs,Hogan_GW_1986,Kosowsky:1992vn,Kosowsky:1992rz,Kamionkowski:1993fg}. For this reason, and in the context of several future GW experiments, it has become crucial to quantify the amplitude and the spectrum of GWs emitted during cosmological PTs. One of the strongest sources of GWs are the sound waves propagating in the plasma after the end of the transition~\cite{Caprini:2019egz, Hindmarsh:2015qta, Hindmarsh:2017gnf}, which are sourced by the kinetic energy deposited in the plasma. 

In the case of direct PTs, it has been showed that the amplitude of the signal from sound waves is controlled by~\cite{Espinosa:2010hh}
\bea 
\Omega^{\rm direct}_{\rm GW} \propto  \bigg(\frac{\rho_{\rm kin}}{\rho_{\rm tot}}\bigg)^2 = \bigg(\frac{\kappa_{\rm direct} \alpha_N}{1+ \alpha_N} \bigg)^2\ , 
\eea 
where $\Omega_{\rm GW}$ is the fraction of energy in GW radiation today, where $\kappa_{\rm direct} \,\alpha_N/(1+ \alpha_N)$ corresponds to the fraction of energy converted to bulk fluid motion. 

In the context of inverse PTs, we expect the GW signal to analogously scale like 
\bea 
\Omega^{\rm inv}_{\rm GW} \propto  \bigg(\frac{\rho_{\rm kin}}{\rho_{\rm tot}}\bigg)^2 = \kappa_{\rm inverse}^2\ .
\eea 
We however emphasise that such claim should be confirmed by numerical simulations.

\section{The pressure on the bubble wall}
\label{sec:friction_inv_PT}

In this section we investigate the pressure and the driving force which acts on the bubble walls, first in the direct case and then in the inverse PT. 
\subsection{The driving force in the direct PT}

Despite the difference in pressure, an important phase transition parameter which cannot in principle be fixed by hydrodynamics only, is the velocity of the bubble wall expansion, $\xi_w$. To investigate it we need to rely on a microphysics analysis. The EoM of the scalar field is given by~\cite{Moore:1995si, Moore:1995ua} 
\begin{align}
 \label{eq:eom}
  \Box\phi+\frac{d V(\phi)}{d\phi}+\sum_i\frac{d m^2_i(\phi)}{d\phi}\int \frac{d^3{\bf p}}{(2\pi)^32E_i}\,f_i(p,x)=0\ ,
 \end{align}
where $f_i$ are the distribution functions of the different particles coupling to the wall. The distribution functions are unknown and need to be solved via Boltzmann equations. 
Following the lines of~\cite{Espinosa:2010hh}, we can integrate the EoM in Eq.\eqref{eq:eom} $\int dz \partial_z \phi$. Upon this operation, we obtain the driving force for the expansion of the bubble\footnote{Notice that there is an intrinsic freedom in determining what is called the \emph{driving force} and what is called the \emph{friction force}. Here we follow the split explained in~\cite{Ai:2024shx}, and used typically in particle physics computation of the pressure~\cite{Bodeker:2009qy} as opposed to~\cite{Espinosa:2010hh}. }
\bea 
F_{\rm vacuum } \equiv \int^{\text{outside}}_{\text{inside}} dz \partial_z \phi \frac{d V(\phi)}{d\phi}  = \epsilon_+ - \epsilon_-\ , 
\eea 
which is positive for a direct phase transition and negative for an inverse phase transition, and the resisting friction originating from the plasma 
\begin{align}
\label{eq:Pfriction_as_a_sum}
    \P_{\rm plasma} \equiv -\int dz \partial_z  \phi \sum_i\frac{d m^2_i(\phi)}{d\phi}\int \frac{d^3{\bf p}}{(2\pi)^32E_i}\,f_i(p,z, T) =\P_{\rm LTE}+\P_{\rm dissipative}\, .
\end{align}
Using a separation of the form $f_i(p, T, z) = f^{\rm eq}(p, T, z) + \delta f_i(p, T, z)$, the friction has been conventionally split into a $\rm LTE$ (local thermal equilibrium) contribution and dissipative contribution. From now on, we will follow the following convention: positive pressure will contribute to the acceleration of the bubble wall while negative pressure will resist it. 

    The dissipative force originates from departure from equilibrium piece $\delta f_i(p, T, z)$, while the LTE originates from heating effects and the $f^{\rm eq}(p, T, z)$ piece. The LTE contribution can be further separated in
\bea 
\P_{\rm LTE} = - \int_{\text{inside}}^{\text{outside}}  dz \partial_z \phi \frac{dV_T (\phi, T)}{d\phi} = - \Delta V_T +\int_{\text{inside}}^{\text{outside}}  dz\frac{\partial V_T}{\partial T} \frac{\partial T}{\partial z} \, .
\eea 
Finally, the pressure budget on the wall is  given by 
\bea 
\label{eq:driving_f}
F_{\rm vacuum} - \P_{\rm plasma} = \epsilon_+ - \epsilon_-  + \Delta V_T -\int_{\text{inside}}^{\text{outside}} dz\frac{\partial V_T}{\partial T} \frac{\partial T}{\partial z} - \mathcal{P}_{\rm dissipative} \,. 
\eea 
Let us first neglect the contribution of the dissipative forces. 
In the context of the direct phase transition $\epsilon_-=0$ and we define $\epsilon_+\equiv \epsilon$, so that Eq.\eqref{eq:driving_f} becomes
\bea 
F_{\rm vacuum} - \P_{\rm plasma} =   \epsilon  + \Delta V_T -\int^{\infty}_{-\infty} dz\frac{\partial V_T}{\partial T} \frac{\partial T}{\partial z} \ ,
\eea 
where the integral is to be performed from the \emph{inside} of the bubble to the \emph{outside} of it. 
We cannot solve exactly the integral without a microscopic description of the phase transition and the change of d.o.f. As a first approximation, we can assume that the phase transition is weak enough so that $T_+ \approx T_-$\footnote{Notice that the strict equality $T_+ = T_-$ is technically not consistent with the LTE assumption, which imposes the conservation of entropy current and thus the saturation of Eq.\,\eqref{eq:increase_of_entropy}.} obtaining
\bea 
F_{\rm vacuum} - \P_{\rm plasma} \approx   \epsilon - \frac{1}{3}\Delta a T_+^4\ ,
\eea 
which we can express in the equivalent form 
\bea 
\label{eq:driv_direct}
F_{\rm vacuum} - \P_{\rm plasma} \approx \frac{3w_+}{4}\bigg(\alpha_+ - \frac{1}{4}(1-b)\bigg)\ ,  \qquad b \equiv a_-/a_+\ . 
\eea 
with $b \leq  1$ being the ratio of the number of relativistic d.o.f inside and outside the bubble. The driving force is maximized when $b =1$ (this remains true even if there is a large heating in front of the wall and $T_+ \gg T_N$). A more precise expression can be obtained within the purely local thermal equilibrium approximation by imposing the conservation of entropy across the reaction front, as was followed in~\cite{Ai:2021kak, Ai:2023see, Ai:2024shx}. Our simplified picture in \eqref{eq:driv_direct} captures nevertheless the physics we are interested in.

We observe that in the instance of $\epsilon \to \frac{1}{3}\Delta a T_+^4$, the driving force vanishes. This is an example of the \emph{hydrodynamic obstruction} discussed for in details in~\cite{Konstandin:2010dm, Balaji:2020yrx,Ai:2021kak, Ai:2023see, Ai:2024shx, Sanchez-Garitaonandia:2023zqz,Krajewski:2024gma}. Intuitively, from a purely hydrodynamic point of view, this can be understood from the presence of a shock wave in front of the wall, which heats up the plasma. It has been showed that this hydrodynamic obstruction is maximal at the Jouguet velocity, which is the crossing between the hybrid and the detonation regime. From Fig.\,\ref{fig:quantities} we observe that $\alpha_+$ has a minimum at the Jouguet velocity, which turns into a weaker driving force in Eq.\,\eqref{eq:driv_direct}.

The dissipative force $\mathcal{P}_{\rm dissipative}$ can be estimated within a particle physics model. It originates from 
\bea 
\mathcal{P}_{\rm dissipative} =  \int dz \partial_z  \phi \sum_i\frac{d m^2_i(\phi)}{d\phi}\int \frac{d^3{\bf p}}{(2\pi)^32E_i}\, (f_i(p,z, T) - f^{\rm eq}_i(p,z, T) ) \ .
\eea 
It however requires a careful solving of the Boltzmann equations to be accounted for~\cite{Moore:1995ua, Moore:1995si,Laurent:2022jrs, Laurent:2020gpg, DeCurtis:2022hlx, DeCurtis:2023hil,  DeCurtis:2024hvh}.

\subsubsection{The runaway solution for direct PTs}

In the regime of fast bubbles, the wall can reach a terminal velocity, described by a boosted detonation, or keep accelerating until the collision. We call the latter, a runaway wall. This type of runaway wall can be studied neglecting the collisions among particles in the wall~\cite{Dine:1992wr,Mancha:2020fzw,Bodeker:2009qy, Bodeker:2017cim,  Vanvlasselaer:2020niz, Gouttenoire:2021kjv, Azatov:2023xem, Ai:2023suz}, where it is assumed that the distribution is fixed all along the wall by the incoming flux: $f_i(p,T) = f_{\rm outside}(p,T_N)$. This avoids the usual split into an equilibrium and an out-of-equilibrium piece, as performed in the previous subsection. This is approximately verified when particles entering do not scatter inside the wall. In the regime of fast bubbles, $\gamma_w \gg 1$, where we defined $\gamma_{w}\equiv 1/\sqrt{1-\xi^2_w}$ the boost factor of the wall to the plasma, it is known that while the contribution from LTE decreases, the most important contribution originates from the particles coupling to the scalar field and thus gaining a mass~\cite{Bodeker:2009qy}. In the case of the fast walls, it has been obtained
\bea 
\label{eq:directP}
\P^{{\rm LO},\gamma_w \to \infty}_{\rm plasma}  \approx \sum_i c_i g_i\frac{\Delta m^2_i T^2}{24} \ , 
\eea  
where $\Delta m^2\equiv m_b^2- m_s^2> 0$ and where $c_i = 1(1/2)$ for bosons (fermions), and $g_i$ is the number of d.of of each particle.  On top of those contributions, pressure from particle splittings studied in~\cite{Bodeker:2017cim,  Vanvlasselaer:2020niz, Gouttenoire:2021kjv, Azatov:2023xem} has been also shown to be able to stop the acceleration of the bubble wall. Notice that such contributions are by definition \emph{not} accounted in Eq.\,\eqref{eq:eom}. This is an open question of how to express particle splittings in the framework of Eq.\,\eqref{eq:eom}. 

 In a given model the condition for runaway behavior is usually formulated as 
\bea 
|\mathcal{P}^{\rm max}_{\rm plasma}(T_{\rm nuc})| < \epsilon \, ,
\eea
where $\mathcal{P}^{\rm max}_{\rm plasma}(T_{\rm nuc})$ contains several pieces: the LO plasma pressure dominated by the expression in Eq.\eqref{eq:directP}, possibly pressure from mixing particles \cite{Azatov:2020nbe} and pressure from emission of soft gauge bosons, scaling like the wall boost factor $\gamma_w$.

\subsection{The driving force in the inverse PT}

In the case of an inverse phase transition, the difference of vacuum energy becomes a resisting pressure and the driving force originates from the plasma effects. We reinterpret the various contributions in the following way: the resisting force is given by
\bea 
F_{\rm vacuum} \equiv \int dz \partial_z \phi \frac{d V(\phi)}{d\phi} = \epsilon_+ - \epsilon_-\ ,  
\eea 
and the pushing plasma effect is given by 
\begin{align}
\label{eq:Pfriction_as_a_sum}
    \P_{\rm plasma} \equiv -\int dz \partial_z  \phi \sum_i\frac{d m^2_i(\phi)}{d\phi}\int \frac{d^3{\bf p}}{(2\pi)^32E_i}\,f_i(p,z, T) =\P_{\rm LTE}+\P_{\rm dissipative}  \ .
\end{align}

As before, we first ignore the dissipative contributions. In the inverse PT case $\epsilon_+=0$, and we define $\epsilon_-\equiv \epsilon$. The approximate LTE expression becomes  (when we can approximate $T_+ \approx T_-$):
\bea 
\label{eq:pressure_budget_inv}
F_{\rm vacuum} - \mathcal{P}_{\rm plasma}  \approx \frac{3w_+}{4}\bigg( \frac{1}{4}(b-1) - |\alpha_+|\bigg)\ ,  \quad b \equiv a_-/a_+ \, , 
\eea 
where now $b> 1$. We observe that the driving force fuelling the expansion now originates from the change of d.o.f. and has to overcome the resisting force from the vacuum. This requires $b > 1+ 4|\alpha_+|$. This approach misses potentially important physical effects, as already stated in the previous section, as for example effects coming from the change of temperature across the wall. We leave this for further studies. 

Looking back at Fig.\,\ref{fig:quantities}, we observe that even at constant $\alpha_N$, $\alpha_+$ is a function of the wall velocity $\xi_w$ and presents a peak at the crossing between the inverse detonation and the inverse hybrid profile.
Notice that, since in the case of inverse transitions the pressure associated with $|\alpha_+|$, as observed in the RHS of Eq.\,\eqref{eq:pressure_budget_inv}, is resisting the expansion, this behavior implies a hydrodynamical obstruction very similar to the one that has been largely studied in direct phase transitions~\cite{Konstandin:2010dm, Ai:2021kak, Ai:2024shx} and recalled in the subsection above. Like in the direct case, the obstruction shows a peak at the (inverse) Jouguet velocity. This trend would deserve more investigation, we however leave the study of the subtleties of such effects to future studies.

\subsubsection{The runaway solution for inverse PTs}
\label{sec:runaway_inv_PT}

So far, we have studied the steady states of the expansion modes. A steady state is reached when the vacuum force is balanced by the plasma pressure. It remains the possibility that a steady state is not reached at all and the bubble keeps accelerating until the collision. 


\emph{What about the possibility of the runaway solution for the inverse PT?} In principle, this can be studied in the collisionless limit, since we consider $\gamma_w \gg 1$: the pressure from the exchange of momentum originates from some particles losing their mass and inducing a kick on the wall. Let us sketch the analysis in the collisionless regime. Since the particles lose their mass when entering the wall, we expect the reflection from outside to be negligible. 
Let us then as a first approximation only consider the entering species: in the wall frame, we can apply the conservation of energy along the particle trajectory,
\bea 
\label{Eq:kick_inverse}
E = \sqrt{m^2 + p^2_z + p_\perp^2}\, , \qquad \frac{dE}{dz} =  \bigg(\frac{dm^2}{dz} + \frac{dp_z^2}{dz}\bigg) \frac{1}{2E} = 0\, ,
\qquad  
\Rightarrow \Delta p^{\rm part}_z \approx -\frac{|\Delta m^2|}{2 p_z}\ .
\eea   
From Eq.\eqref{Eq:kick_inverse}, we see that particles entering the wall and losing their mass are accelerated \emph{inward} by a negative $\Delta p_z$. By conservation of momentum, the wall receives an equal and opposite kick, $ \Delta p^{\rm part}_z = -  \Delta p^{\rm wall}_z > 0 $, which accelerates it forward. This leads us to conclude that the wall is \emph{aspired} as the particles lose their mass. This behavior was dubbed \emph{anti friction} in the analysis of Ref.\,\cite{Buen-Abad:2023hex}.

We need however to convolute with the incoming flux to have a pressure 
\begin{align}
\mathcal{P}_{\rm plasma}
&= -\int dz \partial_z  \phi \sum_i g_i\frac{d m^2_i(\phi)}{d\phi}\int \frac{d^3{\bf p}}{(2\pi)^32E_i}\,f_i(p,z, T) 
\nn 
&\approx - \sum_{i}g_i\int \frac{d^3 p}{(2\pi)^3}  \frac{|\Delta m^2_i|}{2 E_i} f^{\rm eq}_{\rm outside}(p, T) 
\end{align}
where the sum is to be performed over particles losing their mass and $g_i$ is the number of d.o.f of each particle. The integral over the phase space is frame-independent and we compute it in the plasma frame. We obtain the final expression
\bea 
\mathcal{P}^{\gamma_w \to \infty}_{\rm plasma}  \approx  -\sum_{i} C_{\rm eff,i}(m^i_{\rm out}/T) c_i g_i\frac{|\Delta m_i^2| T^2}{24}  \,, 
\eea 
where again $c_i = 1(1/2)$ for bosons (fermions).
We also defined $C_{\rm eff} \leq 1$ to take into account the Boltzmann suppression of the particles outside the bubble. It has an analytical expression in two limits
\bea 
 C_{\rm eff}(m^i_{\rm out}/T) \frac{T^2}{24} \equiv\int\frac{d^3p}{(2\pi)^3 2E_i}  f_i (p, T) \approx  \begin{cases}
\frac{T^2}{24}\quad \text{if} \quad m_i^{\rm out} \ll T\;,
\\
\frac{1}{2 m_i^{\rm out }}\bigg(\frac{m_i^{\rm out} T_{\text{nuc}}}{2\pi}\bigg)^{3/2}e^{-m_i^{\rm out}/T}\quad \text{if} \quad   m_i^{\rm out}\gg T\;.
\end{cases}  
\label{eq:ceff}
\eea 
 and we emphasize that the particles with large masses outside of the bubble become Boltzmann suppressed. 
Therefore, if the following inequality is satisfied,
\bea 
\label{eq:runawayinverse}
 \epsilon < \bigg|\sum_{i} C_{\rm eff,i}(m^i_{\rm out}/T) c_i g_i\frac{|\Delta m_i^2| T^2}{24} \bigg| \, ,
\eea
then the wall can in principle runaway. 

However, we will expect, in the same case as in the direct phase transition~\cite{Azatov:2023xem}, that there might be other sources of pressure from the splitting of particles passing through the wall~\cite{Azatov:2024auq}. 

Notice that the presence of the Boltzmann suppression factor in \eqref{eq:runawayinverse} implies that the BM pressure for inverse and direct transitions can actually be different (in absolute value) for the cooling and the heating phase transition happening in the very same particle physics model. This could change the conclusion in Ref.\,\cite{Buen-Abad:2023hex} that runaway for the direct PT would imply non--runaway for the inverse PT, and viceversa.

Let us however notice that from the point of view of hydrodynamics, the situation for inverse phase transitions appears different than in the case of direct phase transitions. 
In the case of direct phase transitions the hydrodynamical solutions with $\xi_w \sim 1$ corresponding to ultra--relativistic detonations are in principle possible for any value of $\alpha_+ = \alpha_N$. In an inverse phase transition steady--wall solutions with $\xi_w \sim 1$ correspond to inverse deflagrations. These solutions exist only for $-1/3< \alpha_+ <0$. It is then conceivable that when $\alpha_+$ becomes too negative the runaway behavior is prevented.
This reflects the fact that the vacuum energy is now opposed to the bubble expansion rather than fueling it. On the other hand, runaway walls are not steady-state solutions, so this reasoning cannot firmly invalidate runaway for $\alpha_+ < -1/3$.


\section{Conclusion}
\label{sec:conclusion}

\begin{figure}
    \centering
    \includegraphics[width=.33\textwidth]{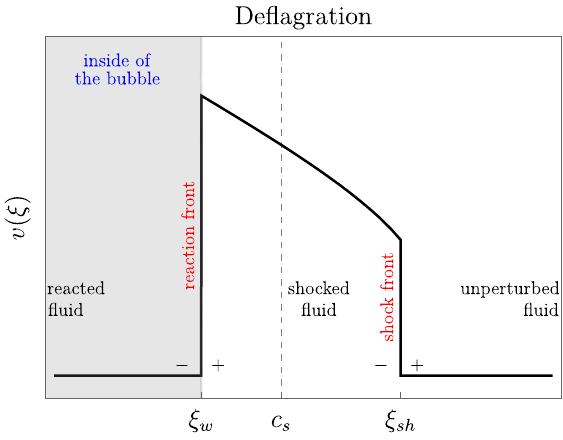}\includegraphics[width=.33\textwidth]{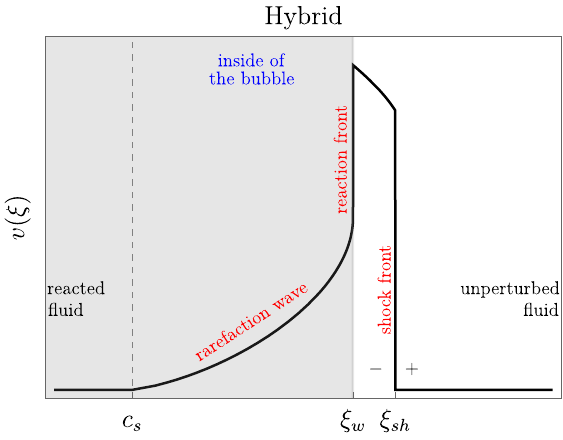}\includegraphics[width=.33\textwidth]{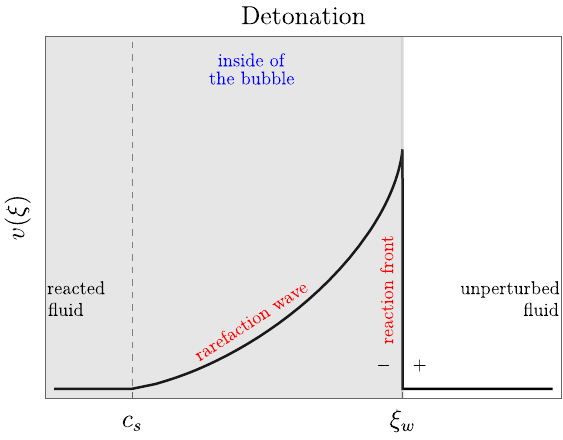}\\
     \includegraphics[width=.33\textwidth]{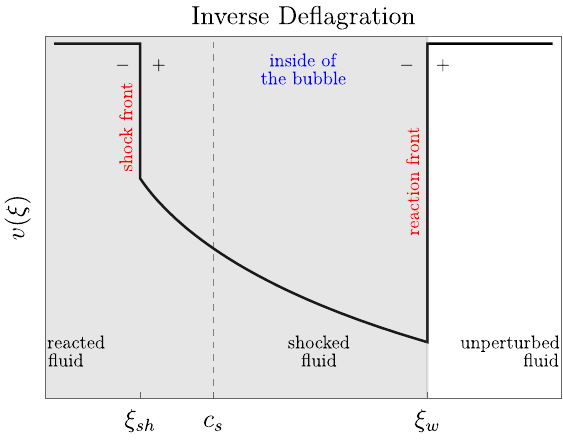}\includegraphics[width=.33\textwidth]{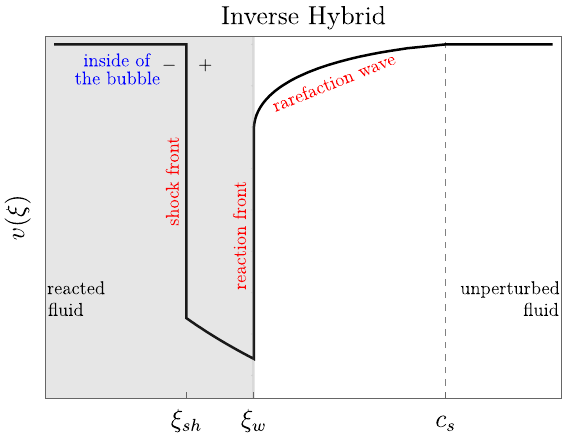}\includegraphics[width=.33\textwidth]{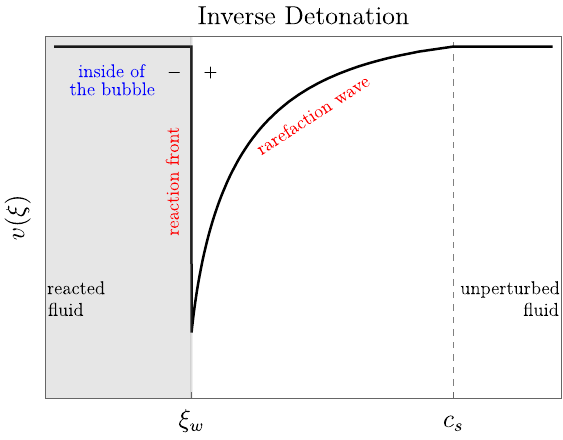}    
    \caption{Cartoon illustrating the various fluid solutions, and the corresponding discontinuities and interfaces for direct PTs (top) and inverse PTs (bottom).
}
    \label{fig:sketch_sol}
\end{figure}

The reheating of the Universe, for example at the end of cosmic inflation~\cite{Buen-Abad:2023hex}, or after a period of matter domination, might lead to inverse phase transitions against the vacuum energy of the zero--temperature potential. Formally, we find that these transitions can be studied by taking $\alpha_N \to -\alpha_N$ in the expression for the fluid velocity as computed from the matching conditions in the case of direct phase transitions. 

In this paper, we have focused on the hydrodynamics of inverse transitions, and we have identified five different modes of bubble expansion: weak and CJ inverse detonations, weak and CJ inverse deflagrations and finally inverse hybrids. We have excluded strong inverse detonations as they are inconsistent with the boundary conditions of the bubble, as well as strong inverse deflagrations because they are unstable and likely excluded by entropy considerations. We show schematically a summary of the possible fluid velocity profiles for inverse PTs in Fig.\,\ref{fig:sketch_sol}, alongside with the direct PT profiles for comparison.   

We have provided the velocity, enthalpy, and temperature profiles for each of these solutions and examined the efficiency for transferring energy into bulk fluid motion after nucleation. This can be used to assess the amplitude of the gravitational waves produced during such a transition. 

Finally, we have studied the pressure exerted on the bubble wall of the inverse phase transition. As opposed to the direct case, the driving force originates from the plasma rather than from the vacuum energy. We also find an analogous hydrodynamic obstruction, where the resistance to the expansion is maximal at the crossing of the inverse Jouguet velocity.

While our study clearly has applications in the case of heating phase transitions, we leave the exploration of particle physics models that could lead to the realization of such inverse phase transitions for future studies.

\section*{Acknowledgements}
 It is a pleasure to thank Wenyuan Ai, Aleksandr Azatov, Jose R. Espinosa, Thomas Konstandin, Benoit Laurent, Rose Leyens, Alberto Mariotti,  Oleksii Matsedonskyi, Jose M. No, Enrico Perboni, Rudin Petrossian-Byrne, Luciano Rezzolla, Geraldine Servant, and Jorinde Van De Vis for helpful discussions and comments on the draft.

SB is supported by the Deutsche Forschungsgemeinschaft under Germany’s Excellence Strategy - EXC 2121 Quantum Universe - 390833306, by FWO-Vlaanderen through grant numbers 12B2323N, and in part by the Strategic Research Program High-Energy Physics of the Research Council
of the Vrije Universiteit Brussel and by the iBOF ``Unlocking the Dark Universe with Gravitational Wave Observations: from Quantum Optics to Quantum Gravity'' of the Vlaamse Interuniversitaire Raad. 
MV is supported by the ``Excellence of Science - EOS" - be.h project n.30820817, and by the Strategic Research Program High-Energy Physics of the Vrije Universiteit Brussel. GB is in part  supported by the MIUR contract 2017L5W2PT.

\appendix

\section{Stability of the hydrodynamical solutions}
\label{app:stability}

The stability of the direct phase transitions has been thoroughly studied in previous works~\cite{bookLandau} and~\cite{RezBook} (sections 4.8 and 5.5) and more specifically in~\cite{Megevand:2014yua,Megevand:2013yua}. In this context, the stability of discontinuities has been related to the concept of \emph{evolutionarity} of the front. A front is said to be evolutionary if any type of infinitesimal perturbations acting on it remains infinitesimal. 

To determine if a front is \emph{evolutionary}, it is sufficient to compute its degree of under-determinacy $D_U$~\cite{bookLandau,RezBook}, which is defined as the difference between the unknown parameters associated with the front and the number of boundary conditions applied on the front. 

The boundary conditions applied on the wall are the three conservation equations, the conservation of mass, momentum, and energy across the front and possibly a fourth condition imposing the front velocity $\xi$. On the other hand, the number of unknown parameters is given by i) the number of acoustic $\lambda^{\pm}_{1,2}$ and 2) the entropy perturbations $\lambda^{\pm}_{0}$ \emph{that can be transmitted} from the front~\cite{bookLandau}. We call those perturbations \emph{propagating}.  
We can obtain that
\begin{itemize}
    \item  The perturbations $\lambda^{\pm}_{1,2}$ in the front frame have the form
\bea 
\text{Ahead}: \qquad \lambda^{+}_{1} = \frac{v_++c_s}{1+v_+ c_s} \propto v_++c_s\qquad \lambda^{+}_{2} = \frac{v_+-c_s}{1-v_+ c_s} \propto v_+-c_s \, ,
\\
\text{Behind}: \qquad \lambda^{-}_{1} = \frac{v_-+c_s}{1+v_- c_s} \propto v_-+c_s\qquad \lambda^{-}_{2} = \frac{v_--c_s}{1-v_- c_s} \propto v_--c_s \, ,
\eea 
where the proportionality is valid in the Newtonian limit and is simply the composition between an acoustic perturbation propagating with a velocity $\pm c_s$ in a fluid with bulk velocity $v_-$ downstream or $v_+$ upstream. A propagating perturbation (for a \emph{left} propagating front) ahead of the wall needs to have $\lambda^{+} <0$ while a propagating perturbation behind the wall will have $\lambda^{-} >0$. Those are the conditions for the perturbations to be propagated away from the front. Perturbations violating those conditions remain stuck on the front and do not propagate.

\item $\lambda^{\pm}_{0}$: On the other hand, the entropy perturbations $\lambda^{\pm}_{0}$ are always transmitted~\cite{bookLandau,RezBook}. The count is thus always two from the entropy perturbations.  
\end{itemize}
Let us illustrate the computation with the case of $v_+ > c_s, v_- < c_s$, we have 
\bea 
\lambda^{+}_{1} > 0  \qquad \lambda^{+}_{2} > 0 
\qquad \lambda^{-}_{1} > 0 \qquad 
 \lambda^{-}_{2} < 0  \, ,
\eea 
where we can conclude that only $\lambda^{-}_{1}$ is propagating. Put in a more intuitive way, in the region where the flow is upstream (toward) the front, $v_+ > c_s$, and so the two acoustic perturbations which propagate with velocity $c_s$ cannot propagate away from the front, since $v_+ > c_s$. On the other hand, downstream, the flow goes out from the wall with velocity $v_- < c_s$ and so, while the perturbation propagating toward the front cannot escape from it, the perturbation propagating away from the front can escape it because $v_- < c_s$. This leads us to the conclusion that for $v_+ > c_s, v_- < c_s$, there are three propagating disturbances, two entropy and one acoustic.
 If the velocity of the shock is not imposed, the degree of under-determinacy is finally $D^{v_+ > c_s, v_- < c_s}_U=3-3 =0$, and the front is \emph{evolutionary}. Following a very similar computation, 
\bea 
D^{v_+ > c_s, v_- < c_s}_U=3-3 =0, \quad D^{v_+ > c_s, v_- > c_s}_U = 1,\quad D^{v_+ < c_s, v_- < c_s}_U = 1, \quad D^{v_+ < c_s, v_- > c_s}_U = 2 \,,
\eea 
\emph{assuming that the velocity of the front is not imposed}. This implies that only a shock wave with $v_+ > c_s, v_- < c_s$ is evolutionary. 

\subsection{Application to direct PTs}
The same analysis can be followed to study the \emph{reaction fronts} like detonations and deflagrations. There is however one important physical difference: while the shock wave velocity cannot be fixed as a boundary condition, the velocity of a phase transition boundary is controlled by hydrodynamics and particle physics. This consists of one more boundary condition. The very same analysis leads to 
\bea 
\label{eq:reaction_stab}
D^{v_+ > c_s, v_- < c_s}_U=3-4 =-1, \quad D^{v_+ > c_s, v_- > c_s}_U = 0,\quad D^{v_+ < c_s, v_- < c_s}_U = 0, \quad D^{v_+ < c_s, v_- > c_s}_U = 1. 
\eea 
The case $v_+ > c_s, v_- < c_s$ corresponds to strong detonations that are not realized in a cosmological phase transition, $v_+ > c_s, v_- > c_s$ and $v_+ < c_s, v_- < c_s$ correspond respectively to weak detonations and weak deflagrations and are evolutionary. Finally, $v_+ < c_s, v_- > c_s$ corresponds to the strong deflagrations and are \emph{not} evolutionary, because even fixing the velocity of the wall does not set the degree of under-determinacy to zero. This analysis has been confirmed by numerical simulations~\cite{Ignatius:1993qn}. 

\subsection{Application to inverse PTs}
We now turn to the stability of the solutions for inverse phase transitions that we discussed in the main text. The results in Eq.\eqref{eq:reaction_stab} remain valid, as they apply directly to the phase boundary, which have the same structure for the direct and the inverse transitions.  

We can thus conclude in exactly the same way: the case $v_+ > c_s, v_- < c_s$ corresponds to the inverse strong detonations,  $v_+ < c_s, v_- < c_s$ and $v_+ > c_s, v_- > c_s$  correspond respectively to weak inverse detonations and weak inverse deflagrations and are evolutionary, and finally $v_+ < c_s, v_- > c_s$ corresponds to strong inverse deflagrations and are \emph{not} evolutionary.

We conclude that direct and inverse strong deflagrations are not evolutionary and then very likely unstable. If they can be made to exist initially, they will split into the hybrid solutions identified in the main text. 

\section{Profiles of the thermodynamic quantities across the waves}
\label{app:thermo}
In this appendix, we collect all the necessary results to compute the plots of Fig.\ref{fig:enthalpy profiles} and the profiles of temperature and enthalpy. 

\subsection{Direct PTs}
We begin by studying the direct phase transitions, with deflagrations, hybrids and detonations:
\paragraph{Deflagrations} Here there are two discontinuities, at the shock and at the wall position, as we can see in Fig. \ref{fig:sketch_sol}. The matching conditions at the shock position are again
\bea 
\label{eq:matching_s}
\text{(matching at the shock)} \qquad w^{sh}_+v^{sh}_+ \gamma^2_+= w^{sh}_- v^{sh}_- \gamma^2_- \ ,
\eea 
where $+$ refers to the right region wrt the shock, i.e. the unperturbed fluid, and -- refers to the left region wrt the shock, the perturbed region. Then we can identify
\bea
\label{eq:shock_V}
w^{sh}_+=w_N \ , \quad v^{sh}_+=\xi_{sh}\ , \quad v^{sh}_-= \mu(\xi_{sh}, v(\xi_{sh}^-))\,  .
\eea 
To obtain the enthalpy after the shock, $w^{sh}_-$, we can plug Eq.\eqref{eq:shock_V} into Eq.\eqref{eq:matching_s} to obtain
\bea 
w^{sh}_- \equiv w(\xi_{sh}^-) = w_N \cdot {\xi_{sh} \over 1-\xi_{sh}^2} \cdot {1-\mu(\xi_{sh}, v(\xi_{sh}^-))^2 \over \mu(\xi_{sh}, v(\xi_{sh}^-))}\ .
\eea 
Now, evolving the solution back to the wall position we have that
\bea 
w(\xi_w^+)=w(\xi_{sh}^-) \cdot \exp \l[ \int_{v(\xi_{sh}^-)}^{v(\xi_{w}^+)}\ \l( 1+ {1 \over c_s^2}\r) \gamma (v)^2 \mu(\xi(v), v) \ dv\r] \ .
\eea 
Proceeding in a similar way as above, the matching conditions at the wall position imply
\bea 
w_-\equiv w(\xi_w^-)= w(\xi_w^+) \cdot {\mu(\xi_{w}, v(\xi_{w}^+))\over 1-\mu(\xi_{w}, v(\xi_{w}^+))^2} \cdot {1- \xi_{w}^2 \over \xi_{w}} \ .  
\eea 
Summarizing the total profile is
\begin{equation}
w(\xi)=\begin{cases}
    w_N & \xi\geq \xi_{sh}^+\\
    w(\xi_{sh}^-) \cdot \exp \l[ \displaystyle\int_{v(\xi_{sh}^-)}^{v(\xi)}\ \l( 1+ \dfrac{1}{c_s^2}\r) \gamma (v)^2 \mu(\xi(v), v) \ dv\r] & \xi_w^+\leq \xi \leq \xi_{sh}^-\\
    w(\xi_w^-)&\xi\leq \xi_w^-
\end{cases}  
\end{equation}  
The same can be applied to the temperature profile which is
\begin{equation}
T(\xi)=\begin{cases}
    T_N & \xi\geq \xi_{sh}^+\\
    T(\xi_{sh}^-) \cdot \exp \l[ \displaystyle\int_{v(\xi_{sh}^-)}^{v(\xi)}\  \gamma (v)^2 \mu(\xi(v), v) \ dv\r] & \xi_w^+\leq \xi \leq \xi_{sh}^-\\
    T(\xi_w^-)&\xi\leq \xi_w^-
\end{cases}  
\end{equation}
where 
\bea 
T(\xi_{sh}^-)&=&T_N \l( w(\xi_{sh}^-) \over w_N \r)^{1/4} \ ,\\
T(\xi_w^+)&=&T(\xi_{sh}^-) \ \exp \l[ \displaystyle\int_{v(\xi_{sh}^-)}^{v(\xi_w^+)}\  \gamma (v)^2 \mu(\xi(v), v) \ dv\r]\ ,\\
T(\xi_w^-)&=&T(\xi_w^+)\l( {w(\xi_{w}^-) \over w(\xi_{w}^+)} \cdot{a_+ \over a_-} \r)^{1/4} \ .
\eea 
The resulting profiles are shown in the right panel (first row) of Fig. \ref{fig:enthalpy profiles}.
\paragraph{Detonations} For the detonations, we only have the discontinuity at the wall position. We skip the detailed derivation and give the resulting profile, that for the enthalpy is
\begin{equation}
w(\xi)=\begin{cases}
    w_N & \xi\geq \xi_{w}^+\\
    w(\xi_{w}^-) \cdot \exp \l[ \displaystyle\int_{v(\xi_{w}^-)}^{v(\xi)}\ \l( 1+ \dfrac{1}{c_s^2}\r) \gamma (v)^2 \mu(\xi(v), v) \ dv\r] & c_s\leq \xi \leq \xi_w^-\\
    w(c_s)&\xi\leq c_s
\end{cases}  
\end{equation}
while the temperature is
\begin{equation}
    T(\xi)=\begin{cases}
    T_N & \xi\geq \xi_{w}^+\\
    T(\xi_{w}^-) \cdot \exp \l[ \displaystyle\int_{v(\xi_{w}^-)}^{v(\xi)}\  \gamma (v)^2 \mu(\xi(v), v) \ dv\r] & c_s\leq \xi \leq \xi_{w}^-\\
    T(c_s)&\xi\leq c_s
\end{cases}
\end{equation}
where
\bea
w(\xi_w^-)&=&w_N \cdot {1-\xi_w^2 \over \xi_w}\cdot{\mu(\xi_w, v(\xi_w^-))\over 1-\mu(\xi_w, v(\xi_w^-))^2}\ ,\\
w(c_s)&=& w(\xi_w^-) \ \exp \l[ \displaystyle\int_{v(\xi_{w}^-)}^{v(c_s)}\ \l( 1+ \dfrac{1}{c_s^2}\r) \gamma (v)^2 \mu(\xi(v), v) \ dv\r]\ ,\\ 
T(\xi_w^-)&=&T_N\l({w(\xi_w^-) \over w_N}\cdot {a_+\over a_-} \r)^{1/4}\ , \\
T(c_s)&=& T(\xi_w^-) \ \exp \l[ \displaystyle\int_{v(\xi_{w}^-)}^{v(c_s)}\ \gamma (v)^2 \mu(\xi(v), v) \ dv\r]\ .
\eea
The profiles are shown in the left panel (first row) in Fig. \ref{fig:enthalpy profiles}.
\paragraph{Hybrids} Here we have two discontinuities and we have to glue the two previous solutions in order to have a consistent one. Skipping the details the resulting enthalpy profile is
\begin{equation}
w(\xi)=\begin{cases}
    w_N & \xi\geq \xi_{sh}^+\\
     w(\xi_{sh}^-) \cdot \exp \l[ \displaystyle\int_{v(\xi_{sh}^-)}^{v(\xi)}\ \l( 1+ \dfrac{1}{c_s^2}\r) \gamma (v)^2 \mu(\xi(v), v) \ dv\r] & \xi_w^+\leq \xi \leq \xi_{sh}^-\\
    w(\xi_{w}^-) \cdot \exp \l[ \displaystyle\int_{v(\xi_{w}^-)}^{v(\xi)}\ \l( 1+ \dfrac{1}{c_s^2}\r) \gamma (v)^2 \mu(\xi(v), v) \ dv\r] & c_s\leq \xi \leq \xi_w^-\\
    w(c_s)&\xi\leq c_s
\end{cases}  
\end{equation}
and analogously for the temperature profile. What is worth noticing is that here, in order to build a consistent solution we glued a CJ deflagration with a rarefaction wave, this means that the velocity of the fluid behind the wall, in the wall frame, has to be $v_-=c_s$, and the matching condition reads
\begin{equation}
    w(\xi_w^-)=w(\xi_w^+) \cdot {\mu(\xi_w, v(\xi_w^+))\over 1-\mu(\xi_w, v(\xi_w^+))^2}\cdot {1-c_s^2 \over c_s} \ .
\end{equation}

\subsection{Inverse PTs}

We now turn to inverse phase transitions.

\paragraph{Inverse Deflagrations} Here, as in the standard case, we have two discontinuities, but now the phases $+$ and $-$ are inverted. We remind the reader that even in this case we keep the identification $+$ to be the region to the right of the discontinuity, while $-$ to be the region to the left. Therefore, using the matching condition, at the wall position we have to impose that
\bea 
w(\xi_w^-)=w_N\cdot {1-\mu(\xi_w, v(\xi_w^-))^2 \over \mu(\xi_w, v(\xi_w^-))}\cdot {\xi_w \over 1-\xi_w^2}\ ,
\eea 
and evolving backward to the shock position we get
\bea 
w(\xi_{sh}^+)=w(\xi_w^-) \ \exp \l[ \displaystyle\int_{v(\xi_{w}^-)}^{v(\xi_{sh}^+)}\ \l( 1+ \dfrac{1}{c_s^2}\r) \gamma (v)^2 \mu(\xi(v), v) \ dv\r]\ .
\eea 
Now, at the shock position, we get that
\bea 
 w(\xi_{sh}^-)=w(\xi_{sh}^+)\cdot {1-\xi_{sh}^2 \over \xi_{sh}} \cdot {\mu (\xi_{sh}, v(\xi_{sh}^+)) \over 1-\mu (\xi_{sh}, v(\xi_{sh}^+))^2}\ . 
\eea 
Summarizing, the enthalpy profile is described by
\begin{equation}
w(\xi)=\begin{cases}
    w_N & \xi\geq \xi_{w}^+\\
    w(\xi_{w}^-) \cdot \exp \l[ \displaystyle\int_{v(\xi_{w}^-)}^{v(\xi)}\ \l( 1+ \dfrac{1}{c_s^2}\r) \gamma (v)^2 \mu(\xi(v), v) \ dv\r] & \xi_{sh}^+\leq \xi \leq \xi_{w}^-\\
    w(\xi_{sh}^-)&\xi\leq \xi_{sh}^-
\end{cases}  
\end{equation}  
The same can be applied to the temperature profile which is
\begin{equation}
T(\xi)=\begin{cases}
    T_N & \xi\geq \xi_{w}^+\\
    T(\xi_{w}^-) \cdot \exp \l[ \displaystyle\int_{v(\xi_{w}^-)}^{v(\xi)}\  \gamma (v)^2 \mu(\xi(v), v) \ dv\r] & \xi_{sh}^+\leq \xi \leq \xi_{w}^-\\
    T(\xi_{sh}^-)&\xi\leq \xi_{sh}^-
\end{cases}  
\end{equation}
where 
\bea 
T(\xi_{w}^-)&=&T_N \l( {w(\xi_{w}^-) \over w_N}\cdot {a_+\over a_-} \r)^{1/4} \ ,\\
T(\xi_{sh}^+)&=&T(\xi_{w}^-) \ \exp \l[ \displaystyle\int_{v(\xi_{w}^-)}^{v(\xi_{sh}^+)}\  \gamma (v)^2 \mu(\xi(v), v) \ dv\r]\ ,\\
T(\xi_{sh}^-)&=&T(\xi_{sh}^+)\l( {w(\xi_{sh}^-) \over w(\xi_{sh}^+)}  \r)^{1/4} \ .
\eea 
The resulting profiles are shown in the right panel (second row) of Fig. \ref{fig:enthalpy profiles}.
\paragraph{Inverse Detonations} Here we only have the discontinuity at the wall position. As before, we skip the detailed derivation and give the resulting profile, that the enthalpy is
\begin{equation}
w(\xi)=\begin{cases}
    w_N & \xi\geq c_s\\
    w_N \cdot \exp \l[ \displaystyle\int_{v(c_s)}^{v(\xi)}\ \l( 1+ \dfrac{1}{c_s^2}\r) \gamma (v)^2 \mu(\xi(v), v) \ dv\r] & \xi_w^+\leq \xi \leq c_s\\
    w(\xi_w^-)&\xi\leq \xi_w^-
\end{cases}  
\end{equation}
while the temperature is
\begin{equation}
    T(\xi)=\begin{cases}
    T_N & \xi\geq c_s\\
    T_N \cdot \exp \l[ \displaystyle\int_{v(c_s)}^{v(\xi)}\ \gamma (v)^2 \mu(\xi(v), v) \ dv\r] & \xi_w^+\leq \xi \leq c_s\\
    T(\xi_w^-)&\xi\leq \xi_w^-
\end{cases}
\end{equation}
where
\bea
w(\xi_w^-)&=&w(\xi_w^+) \cdot {1-\xi_w^2 \over \xi_w}\cdot{\mu(\xi_w, v(\xi_w^+))\over 1-\mu(\xi_w, v(\xi_w^+))^2}\ ,\\
w(\xi_w^+)&=& w_N \ \exp \l[ \displaystyle\int_{v(c_s)}^{v(\xi_{w}^+)}\ \l( 1+ \dfrac{1}{c_s^2}\r) \gamma (v)^2 \mu(\xi(v), v) \ dv\r]\ ,\\ 
T(\xi_w^-)&=&T(\xi_w^+)\l({w(\xi_w^-) \over w(\xi_w^+)}\cdot {a_+\over a_-} \r)^{1/4}\ , \\
T(\xi_w^+)&=& T_N \ \exp \l[ \displaystyle\int_{v(c_s)}^{v(\xi_{w}^+)}\ \gamma (v)^2 \mu(\xi(v), v) \ dv\r]\ .
\eea
The profiles are shown in the left panel (second row) in Fig. \ref{fig:enthalpy profiles}.
\paragraph{Inverse Hybrids} Here we have two discontinuities and we have to glue the two previous solutions in order to have a consistent one. Skipping the details the resulting enthalpy profile is
\begin{equation}
w(\xi)=\begin{cases}
    w_N & \xi\geq c_s\\
     w_N \cdot \exp \l[ \displaystyle\int_{v(c_s)}^{v(\xi)}\ \l( 1+ \dfrac{1}{c_s^2}\r) \gamma (v)^2 \mu(\xi(v), v) \ dv\r] & \xi_w^+\leq \xi \leq c_s\\
    w(\xi_{w}^-) \cdot \exp \l[ \displaystyle\int_{v(\xi_{w}^-)}^{v(\xi)}\ \l( 1+ \dfrac{1}{c_s^2}\r) \gamma (v)^2 \mu(\xi(v), v) \ dv\r] & \xi_{sh}^-\leq \xi \leq \xi_w^-\\
    w(\xi_{sh}^-)&\xi\leq \xi_{sh}^-
\end{cases}  
\end{equation}
and analogously for the temperature profile. What is worth noticing, even in the inverse case, is that, in order to build a consistent solution we glued a CJ inverse deflagration with an inverse rarefaction wave, this means that the velocity of the fluid in front of the wall, in the wall 
frame, has to be $v_+=c_s$, and the matching condition reads
\begin{equation}
    w(\xi_w^-)=w(\xi_w^+) \cdot {1-\mu(\xi_w, v(\xi_w^-))^2 \over \mu(\xi_w, v(\xi_w^-))}\cdot { c_s\over 1-c_s^2 } \ .
\end{equation}
The resulting profiles are shown in the middle panel (second row) of Fig.\ref{fig:enthalpy profiles}.

\section{Taub and Poisson's adiabats}
\label{app:taub_and_react_adiabat}

In this section, we introduce how discontinuities are treated within hydrodynamics and how to connect states across such interfaces. We will introduce the Taub adiabat and summarize the main features. First seminal works, to name a few, were done by~\cite{PhysRev.94.1468,1973ApJ...179..897T} , but we will follow the more modern approach presented in~\cite{bookLandau, RezBook}.

In the presence of discontinuity waves, like shock and reaction fronts, the profile can be mathematically described by a discontinuity surface with a region ahead of the shock (upstream) and one behind it (downstream). The discontinuity prevent to use the relativistic hydrodynamic equation across the discontinuity, which can be replaced by the relativistic version of the Rankine-Hugoniot junction conditions. We can recast the hydrodynamic equations
\bea 
\nabla_\mu(\rho u^\mu)=0\ , \qquad \nabla_\mu T^{\mu \nu}=0\ ,
\eea 
in global equation in the sense that they are not anymore only locally defined. To do so we choose an arbitrary function $f$ and a $4-$vector $\lambda_\nu$ and, using the previous expression, we can write
\bea 
\nabla_\mu(\rho u^\mu f)=\rho u^\mu \nabla_\mu f \ , \qquad \nabla_\mu(T^{\mu \nu}\lambda_\nu)=T^{\mu\nu}\nabla_\mu\lambda_\nu \ .
\eea
We then integrate both sides over an arbitrary region $\mathcal{V}$ containing the shock surface and all his history, $\Sigma$, and then apply Stoke's theorem on the LHS, obtaining
\bea
\int_\mathcal{S}\rho u^\mu fn_\mu\ dV&= {\int_\mathcal{V}}\rho u^\mu \nabla_\mu f \ d^4x \ , \\
\int_\mathcal{S}T^{\mu \nu}\lambda_\nu n_\mu \ dV&= \int_\mathcal{V}T^{\mu\nu}\nabla_\mu\lambda_\nu \ d^4x \ ,
\eea 
where $n_\mu$ is the $4-$vector perpendicular to $\mathcal{S}$. We can now shrink $\mathcal{V}$ to zero while comprising a portion $\Sigma'$ of the $3$D worldline relative to the shock front. In this limit, LHS$\to0$ while the RHS is the quantity computed at each side of the shock, then
\bea
\int_{\Sigma'}\ f \llbracket\rho u^\mu\rrbracket  n_\mu \ dV=0 \ , \qquad \int_{\Sigma'}\ \lambda_\mu \llbracket T^{\mu \nu}\rrbracket  n_\mu \ dV=0 \ ,
\eea 
where we introduced the double bracket notation $\llbracket A\rrbracket =A_+-A_-$. Then the conditions
\bea
\llbracket \rho u^\mu\rrbracket  n_\mu=0 \ , \qquad \llbracket T^{\mu \nu}\rrbracket  n_\mu =0 \ ,
\eea 
are exactly the junction conditions written in the main text, in Eqs. \eqref{eq:conditionA} and \eqref{eq:conditionB}. Now, defining the mass flux as $J=\rho \gamma v=const.$, and taking $u^\mu=(0,0,0,1)$, we can recast the first the conditions $\rho u^z=const.$ and $T^{zz}=const.$ in the following expressions
\bea 
\label{eq: braket cond}
\llbracket J^2\rrbracket =0\ , \qquad J^2=- {\llbracket p\rrbracket  \over \llbracket w/\rho^2\rrbracket } .
\eea 
Now, from $T^{tz}=const.$ we get
\bea 
w\gamma^2v=const. \quad \to \quad \llbracket \gamma w / \rho \rrbracket=0 \ .
\eea 
Taking the right expression in eq. \eqref{eq: braket cond} multiply it by ${{w_+\over\rho_+^2}+{w_-\over\rho_-^2}}$, and subtracting from it the square of the previous expression we end up in
\bea
\label{eq: braket cond2}
\l\llbracket {w^2\over \rho^2}\r\rrbracket =\l({w_+\over\rho_+^2}+{w_-\over\rho_-^2}\r) \llbracket p\rrbracket  \ .
\eea 
This expression is the relativistic generalization of the classical Hugoniot adiabat for Newtonian shock fronts. Equations \eqref{eq: braket cond} and \eqref{eq: braket cond2} are known as Taub’s junction conditions for shock waves, serving as the relativistic counterparts to the Rankine–Hugoniot junction conditions for Newtonian shocks.

\begin{figure}
    \centering
    \includegraphics[width=.48\textwidth]{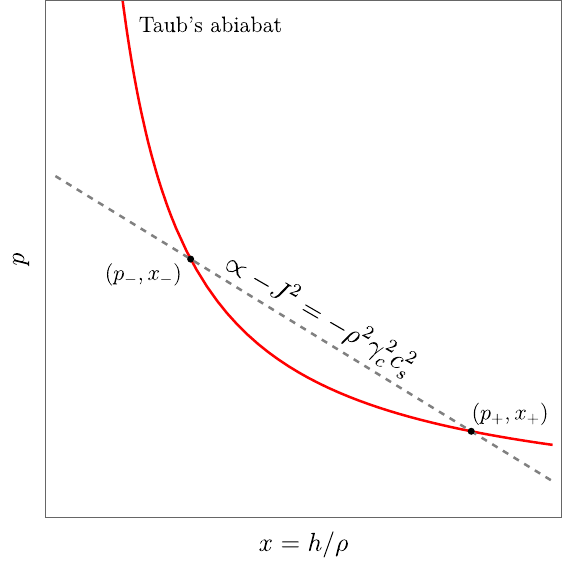}\includegraphics[width=.48\textwidth]{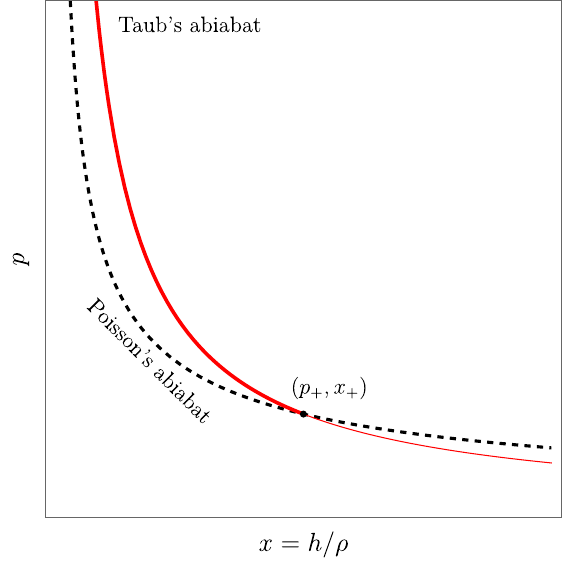}
    \caption{Taub and Poisson's adiabats. In the left panel, it is presented the Taub adiabat connecting two different states, $(x_{\pm}, p_{\pm})$, across a shock, whose chord connecting them is proportional to the (conserved) matter flux $J^2$. In the right panel, the Poisson adiabad is compared to the Taub one. Since this latter is an isoentropic adiabat, i.e. connect states with the same entropy, it selects the physical branch of the Taub adiabat, for which the final state, across the shock, has increased its entropy.}
    \label{fig:Taub}
\end{figure}

When examining the $(p, w/\rho^2)$ plane, the Taub adiabat offers a straightforward and visual representation of fluid properties across a shock. We plotted the shock adiabatic in this plane since the natural variables for representing the relativistic shock adiabatic are $w/\rho^2=wV^2$ and $pc^2$; in these coordinates, $J^2$ gives the slope of the chord from the initial point 1 on the adiabatic to any other point 2. Here, the Taub adiabat appears as the curve connecting the states before and after a shock wave, as illustrated in Fig. \ref{fig:Taub}. Upon defining the state ``+" of the fluid ahead of the front in terms of pressure $p_+$ and the ratio $w_+/\rho^2_+$, condition \eqref{eq: braket cond2} constrains the possible states of the fluid in the shocked region ``$-$" to lie on the adiabat. Similar to the Newtonian result, the chord connecting the two states is proportionate to the square of the mass flux, $J^2$, implying that for the same initial state ``+", a new state ``$-$" with higher pressure will also entail a greater mass flux across the shock.

From the definition of the sound speed
\bea 
c_s^2= \l( d p \over d e \r)_s \ ,
\eea 
and using that for the isoentropic transformation we have that $de=hd\rho$ we can see that
\bea
c_s^2=\l( d p \over d e \r)_s= \l( d p \over d (h/\rho) \r)_s\l( d (h/\rho) \over d e \r)_s=\l( d p \over d (h/\rho) \r)_s {c_s^2-1 \over \rho^2} \ ,
\eea 
that translate in
\bea 
\label{eq: neg derivative Taub}
\l( d p \over d (h/\rho) \r)_s=- {\rho^2 c_s^2 \over 1-c_s^2}=-\rho^2\gamma_s^2 c_s^2 <0 \ ,
\eea 
meaning that the slope at any point along the Taub adiabat is inherently negative and directly proportional to the local sound speed. Consequently, shocked states with higher pressures and densities will exhibit larger sound speeds.

For weak discontinuities\footnote{Discontinuity for which every quantity is small, i.e. the states ahead and behind the shock are not very different.} the jumps in the speciﬁc entropy and in the pressure scale, according to \cite{1973ApJ...179..897T}, as
\bea 
\label{eq: change in entropy}
\llbracket s \rrbracket= \l[ {1 \over 12 h T} \l( d(h/\rho)\over dp\r)_s \r] \llbracket p \rrbracket^3+ O(\llbracket p\rrbracket^4)\ .
\eea 
The meaning of the previous equation lies in the necessity for a change in entropy across a shock, albeit only slightly. Because of the second law of thermodynamics, this change must manifest as an increase. Hence, 
\bea 
s_->s_+ \ .
\eea 
This inequality mirrors the intricate irreversible processes occurring within the narrow region of the shock front. It enables the exclusion of unphysical shocks that fail to induce an increase in entropy.

Let's visualize the situation described: consider the point $(p_+,x_+)$ where $x=h/\rho=w/\rho^2$ in the $p-x$ plane. We draw two curves through this point: the shock Taub adiabatic and the Poisson adiabatic. The equation of the Poisson adiabatic, being an isoentropic transformation, is given by $s_+-s_-= 0$. One can demonstrate that the Poisson and Taub adiabats passing through a given state share identical first and second derivatives at that particular state. Moreover, excluding non-convex equations of state, it can be proven that the second derivative of the Taub adiabat is always positive. To determine the relative position of the two curves near point 1, we consider the fact that we must have $s_- > s_+$ on the shock adiabatic for $p_->p_+$, while on the Poisson adiabatic $s_- = s_+$. Consequently, the abscissa of a point on the shock adiabatic must exceed that of a point on the Poisson adiabatic with the same ordinate $p_-$. When combined, $s_- > s_+$ and $p_->p_+$, they establish that the Poisson adiabat selects the "physical" branch of the Taub adiabat, ensuring it always lies above the constant-entropy curve.

Under the assumption that the concavity of the Taub adiabat is always positive, combined with the increase in pressure across the discontinuity and Eq. \eqref{eq: braket cond2}, we also conclude that $\rho_-w_->\rho_+w_+$. Furthermore, since $w_->w_+$, we deduce $\rho_->\rho_+$.

These inequalities provide crucial insights into the flow velocities on either side of the shock and how they compare with the local sound speed. By using Eq. \eqref{eq:conditionB} and the aforementioned inequalities, we arrive at the conclusion that the velocity must decrease in magnitude across the shock.
\bea 
|v_-|<|v_+| \ .
\eea 
Furthermore, from Eq. \eqref{eq: neg derivative Taub}, along with the previous expression, we can demonstrate that the flow entering a shock front is always supersonic, while the flow exiting it is necessarily subsonic
\bea 
J^2>-\l({\partial p \over \partial x}\r)_+ \quad &\to \quad v_+> c_{s,+}=c_s\ ,\\
J^2<-\l({\partial p \over \partial x}\r)_- \quad &\to \quad v_-< c_{s,-}=c_s\ .
\eea 
It's important to emphasize that the previous inequalities hold true for relativistic (as well as non-relativistic) shock waves, irrespective of the thermodynamic conditions. This is due to the necessity for the shock to be evolutionary, i.e., stable under small perturbations.

\begin{figure}
    \centering
    \includegraphics[width=.48\textwidth]{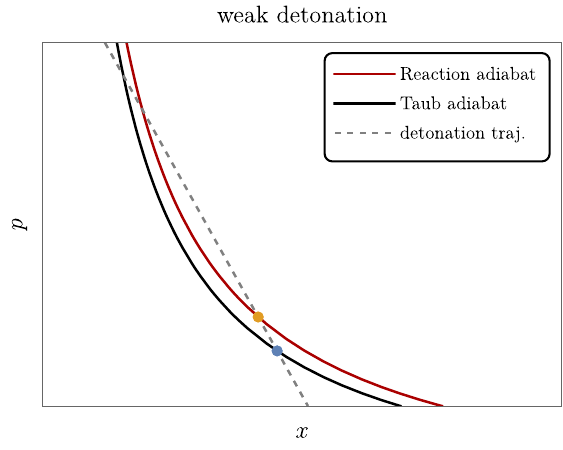}\includegraphics[width=.48\textwidth]{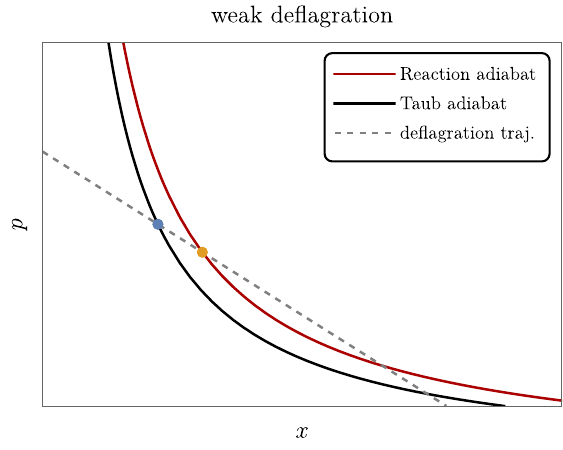}
    \caption{Representation of a weak detonation (left) and a weak deflagration (right). The state of the plasma \emph{ahead} of the wall is represented by the blue dot and lies on the \emph{Taub} adiabat, while the state of the plasma \textit{behind} the discontinuity is represented by the orange dot and lies on the reaction adiabat. }
    \label{fig:whatW}
\end{figure}

\begin{figure}
    \centering
    \includegraphics[width=.48\textwidth]{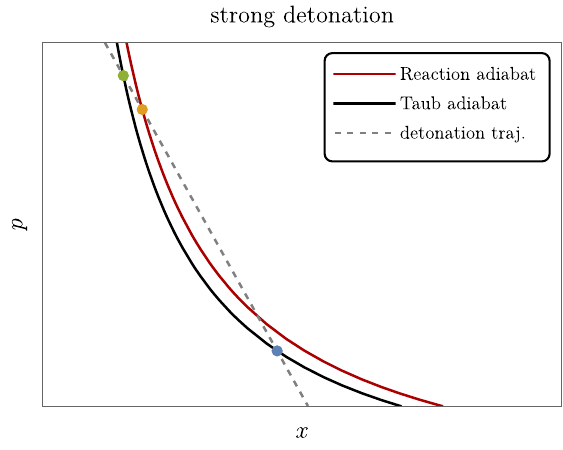}\includegraphics[width=.48\textwidth]{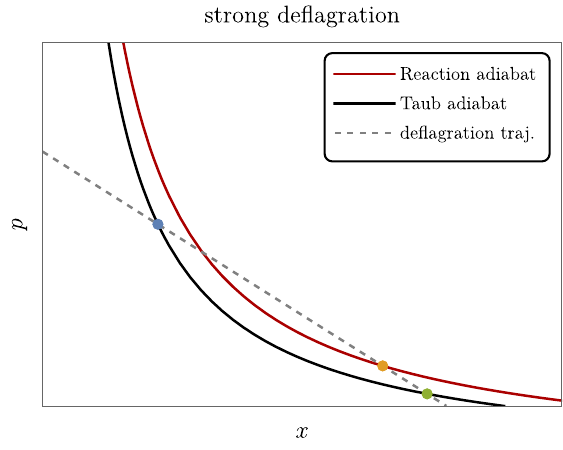}
    \caption{\textbf{Left}: Strong detonation: the physical detonation is the trajectory between the blue point and the orange one (from Taub to the reaction adiabat). However, we can virtually see it as the trajectory from the blue dot to the green one (from Taub to Taub, so a shock wave) and then back to orange (from Taub to the reaction adiabat).
    \textbf{Right}: Strong deflagration: the physical deflagration is the trajectory from the blue dot to the orange one, but, here too, we can see it as the trajectory from the blue to the green and then back to orange.
    }
    \label{fig:what}
\end{figure}

\begin{figure}
    \centering
    \includegraphics[width=.48\textwidth]{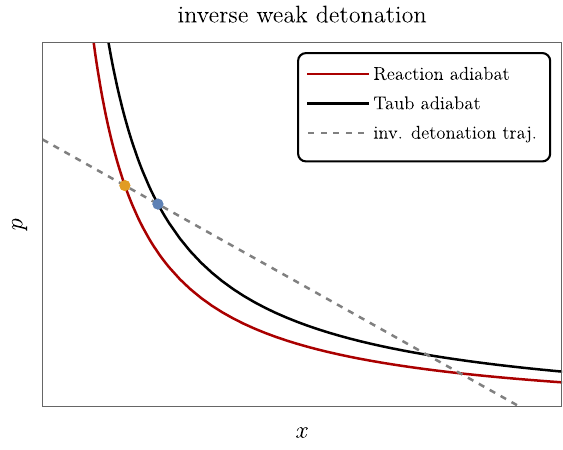}\includegraphics[width=.48\textwidth]{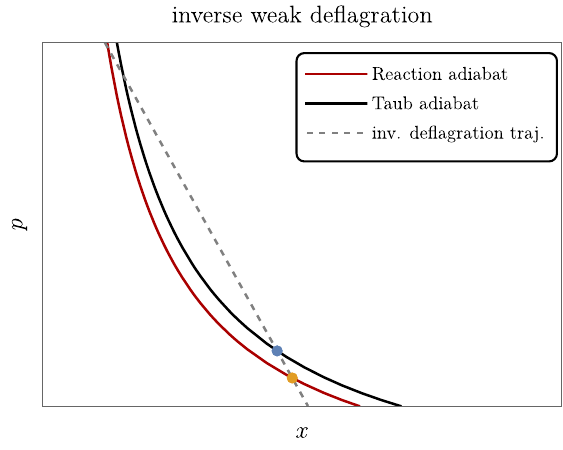}
    \caption{Same as in Fig.\,\ref{fig:whatW} for inverse PT. 
    }
    \label{fig:whatA}
\end{figure}

\begin{figure}
    \centering
    \includegraphics[width=.48\textwidth]{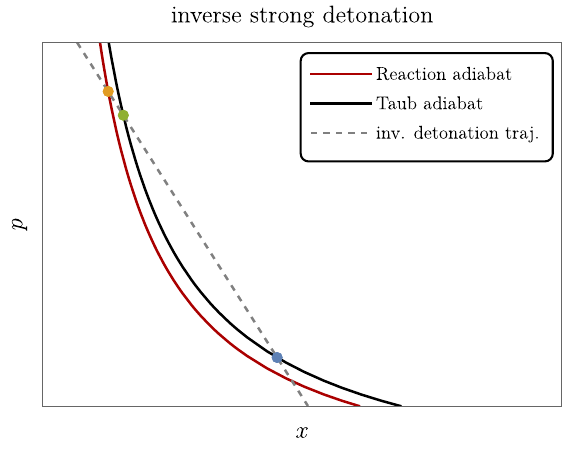}\includegraphics[width=.48\textwidth]{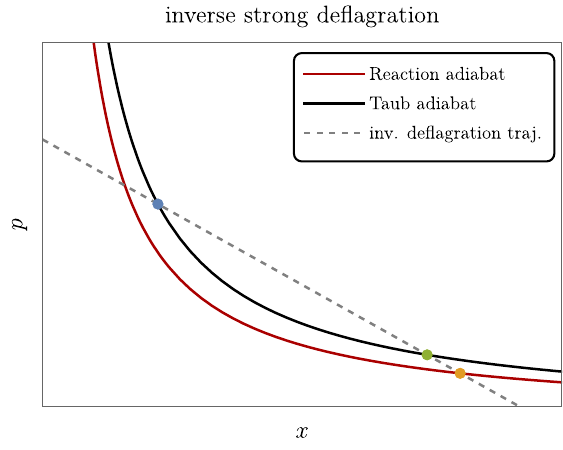}
    \caption{Same as in Fig.\,\ref{fig:what} for inverse PT.  
    }
    \label{fig:whatWA}
\end{figure}

In summary, across a shock, several changes occur: entropy, enthalpy, pressure, and rest-mass density increase, while the velocity relative to the shock decreases. Additionally, the flow is supersonic ahead of the shock and subsonic behind it. Moreover, both the mass flux and the entropy jump increase for states further along the physical branch of the Taub adiabat. These findings hold even for strong shocks, where the jumps in states ahead and behind the shock can be arbitrarily large.
Finally, we here collect all the representations of the Taub and reaction adiabats for the strong and weak, direct and inverse deflagration and detonation in Fig.\ref{fig:whatW}, \ref{fig:what}, \ref{fig:whatWA} and \ref{fig:whatA}.

\bibliographystyle{JHEP}
{\footnotesize
\bibliography{biblio}}

\providecommand{\href}[2]{#2}\begingroup\raggedright\begin{thebibliography}{100}

\bibitem{Kuzmin:1985mm}
V.~A. Kuzmin, V.~A. Rubakov, and M.~E. Shaposhnikov {\em Phys. Lett. B} {\bf
  155} (1985) 36.

\bibitem{Shaposhnikov:1986jp}
M.~Shaposhnikov {\em JETP Lett.} {\bf 44} (1986) 465--468.

\bibitem{Nelson:1991ab}
A.~E. Nelson, D.~B. Kaplan, and A.~G. Cohen {\em Nucl. Phys. B} {\bf 373}
  (1992) 453--478.

\bibitem{Carena:1996wj}
M.~Carena, M.~Quiros, and C.~E.~M. Wagner {\em Phys. Lett. B} {\bf 380} (1996)
  81--91, [\href{http://arxiv.org/abs/hep-ph/9603420}{{\tt hep-ph/9603420}}].

\bibitem{Cline:2017jvp}
J.~M. Cline {\em Phil. Trans. Roy. Soc. Lond. A} {\bf 376} (2018), no.~2114
  20170116, [\href{http://arxiv.org/abs/1704.08911}{{\tt arXiv:1704.08911}}].

\bibitem{Long:2017rdo}
A.~J. Long, A.~Tesi, and L.-T. Wang {\em JHEP} {\bf 10} (2017) 095,
  [\href{http://arxiv.org/abs/1703.04902}{{\tt arXiv:1703.04902}}].

\bibitem{Bruggisser:2018mrt}
S.~Bruggisser, B.~Von~Harling, O.~Matsedonskyi, and G.~Servant {\em JHEP} {\bf
  12} (2018) 099, [\href{http://arxiv.org/abs/1804.07314}{{\tt
  arXiv:1804.07314}}].

\bibitem{Bruggisser:2018mus}
S.~Bruggisser, B.~Von~Harling, O.~Matsedonskyi, and G.~Servant {\em Phys. Rev.
  Lett.} {\bf 121} (2018), no.~13 131801,
  [\href{http://arxiv.org/abs/1803.08546}{{\tt arXiv:1803.08546}}].

\bibitem{Bruggisser:2022rdm}
S.~Bruggisser, B.~von Harling, O.~Matsedonskyi, and G.~Servant {\em JHEP} {\bf
  08} (2023) 012, [\href{http://arxiv.org/abs/2212.11953}{{\tt
  arXiv:2212.11953}}].

\bibitem{Morrissey:2012db}
D.~E. Morrissey and M.~J. Ramsey-Musolf {\em New J. Phys.} {\bf 14} (2012)
  125003, [\href{http://arxiv.org/abs/1206.2942}{{\tt arXiv:1206.2942}}].

\bibitem{Azatov:2021irb}
A.~Azatov, M.~Vanvlasselaer, and W.~Yin {\em JHEP} {\bf 10} (2021) 043,
  [\href{http://arxiv.org/abs/2106.14913}{{\tt arXiv:2106.14913}}].

\bibitem{Huang:2022vkf}
P.~Huang and K.-P. Xie {\em JHEP} {\bf 09} (2022) 052,
  [\href{http://arxiv.org/abs/2206.04691}{{\tt arXiv:2206.04691}}].

\bibitem{Baldes:2021vyz}
I.~Baldes, S.~Blasi, A.~Mariotti, A.~Sevrin, and K.~Turbang {\em Phys. Rev. D}
  {\bf 104} (2021), no.~11 115029, [\href{http://arxiv.org/abs/2106.15602}{{\tt
  arXiv:2106.15602}}].

\bibitem{Chun:2023ezg}
E.~J. Chun, T.~P. Dutka, T.~H. Jung, X.~Nagels, and M.~Vanvlasselaer
  \href{http://arxiv.org/abs/2305.10759}{{\tt arXiv:2305.10759}}.

\bibitem{Falkowski:2012fb}
A.~Falkowski and J.~M. No {\em JHEP} {\bf 02} (2013) 034,
  [\href{http://arxiv.org/abs/1211.5615}{{\tt arXiv:1211.5615}}].

\bibitem{Baldes:2020kam}
I.~Baldes, Y.~Gouttenoire, and F.~Sala {\em JHEP} {\bf 04} (2021) 278,
  [\href{http://arxiv.org/abs/2007.08440}{{\tt arXiv:2007.08440}}].

\bibitem{Hong:2020est}
J.-P. Hong, S.~Jung, and K.-P. Xie {\em Phys. Rev. D} {\bf 102} (2020), no.~7
  075028, [\href{http://arxiv.org/abs/2008.04430}{{\tt arXiv:2008.04430}}].

\bibitem{Azatov:2021ifm}
A.~Azatov, M.~Vanvlasselaer, and W.~Yin {\em JHEP} {\bf 03} (2021) 288,
  [\href{http://arxiv.org/abs/2101.05721}{{\tt arXiv:2101.05721}}].

\bibitem{Baldes:2021aph}
I.~Baldes, Y.~Gouttenoire, F.~Sala, and G.~Servant {\em JHEP} {\bf 07} (2022)
  084, [\href{http://arxiv.org/abs/2110.13926}{{\tt arXiv:2110.13926}}].

\bibitem{Asadi:2021pwo}
P.~Asadi, E.~D. Kramer, E.~Kuflik, G.~W. Ridgway, T.~R. Slatyer, and J.~Smirnov
  {\em Phys. Rev. D} {\bf 104} (2021), no.~9 095013,
  [\href{http://arxiv.org/abs/2103.09827}{{\tt arXiv:2103.09827}}].

\bibitem{Lu:2022paj}
P.~Lu, K.~Kawana, and K.-P. Xie {\em Phys. Rev. D} {\bf 105} (2022), no.~12
  123503, [\href{http://arxiv.org/abs/2202.03439}{{\tt arXiv:2202.03439}}].

\bibitem{Baldes:2022oev}
I.~Baldes, Y.~Gouttenoire, and F.~Sala {\em SciPost Phys.} {\bf 14} (2023) 033,
  [\href{http://arxiv.org/abs/2207.05096}{{\tt arXiv:2207.05096}}].

\bibitem{Azatov:2022tii}
A.~Azatov, G.~Barni, S.~Chakraborty, M.~Vanvlasselaer, and W.~Yin {\em JHEP}
  {\bf 10} (2022) 017, [\href{http://arxiv.org/abs/2207.02230}{{\tt
  arXiv:2207.02230}}].

\bibitem{Baldes:2023fsp}
I.~Baldes, M.~Dichtl, Y.~Gouttenoire, and F.~Sala
  \href{http://arxiv.org/abs/2306.15555}{{\tt arXiv:2306.15555}}.

\bibitem{Kierkla:2022odc}
M.~Kierkla, A.~Karam, and B.~Swiezewska {\em JHEP} {\bf 03} (2023) 007,
  [\href{http://arxiv.org/abs/2210.07075}{{\tt arXiv:2210.07075}}].

\bibitem{Giudice:2024tcp}
G.~F. Giudice, H.~M. Lee, A.~Pomarol, and B.~Shakya
  \href{http://arxiv.org/abs/2403.03252}{{\tt arXiv:2403.03252}}.

\bibitem{10.1143/PTP.68.1979}
H.~Kodama, M.~Sasaki, and K.~Sato {\em Progress of Theoretical Physics} {\bf
  68} (12, 1982) 1979--1998,
  [\href{http://arxiv.org/abs/https://academic.oup.com/ptp/article-pdf/68/6/1979/5311817/68-6-1979.pdf}{{\tt
  https://academic.oup.com/ptp/article-pdf/68/6/1979/5311817/68-6-1979.pdf}}].

\bibitem{Kawana:2021tde}
K.~Kawana and K.-P. Xie {\em Phys. Lett. B} {\bf 824} (2022) 136791,
  [\href{http://arxiv.org/abs/2106.00111}{{\tt arXiv:2106.00111}}].

\bibitem{Jung:2021mku}
T.~H. Jung and T.~Okui \href{http://arxiv.org/abs/2110.04271}{{\tt
  arXiv:2110.04271}}.

\bibitem{Gouttenoire:2023naa}
Y.~Gouttenoire and T.~Volansky \href{http://arxiv.org/abs/2305.04942}{{\tt
  arXiv:2305.04942}}.

\bibitem{Lewicki:2023ioy}
M.~Lewicki, P.~Toczek, and V.~Vaskonen
  \href{http://arxiv.org/abs/2305.04924}{{\tt arXiv:2305.04924}}.

\bibitem{Witten:1984rs}
E.~Witten {\em Phys. Rev.} {\bf D30} (1984) 272--285.

\bibitem{Hogan_GW_1986}
C.~J. Hogan {\em Mon. Not. Roy. Astron. Soc.} {\bf 218} (1986) 629--636.

\bibitem{Kosowsky:1992vn}
A.~Kosowsky and M.~S. Turner {\em Phys. Rev.} {\bf D47} (1993) 4372--4391,
  [\href{http://arxiv.org/abs/astro-ph/9211004}{{\tt astro-ph/9211004}}].

\bibitem{Kosowsky:1992rz}
A.~Kosowsky, M.~S. Turner, and R.~Watkins {\em Phys. Rev. Lett.} {\bf 69}
  (1992) 2026--2029.

\bibitem{Kamionkowski:1993fg}
M.~Kamionkowski, A.~Kosowsky, and M.~S. Turner {\em Phys. Rev.} {\bf D49}
  (1994) 2837--2851, [\href{http://arxiv.org/abs/astro-ph/9310044}{{\tt
  astro-ph/9310044}}].

\bibitem{Pasechnik:2023hwv}
R.~Pasechnik, M.~Reichert, F.~Sannino, and Z.-W. Wang {\em JHEP} {\bf 02}
  (2024) 159, [\href{http://arxiv.org/abs/2309.16755}{{\tt arXiv:2309.16755}}].

\bibitem{Azatov:2020nbe}
A.~Azatov and M.~Vanvlasselaer {\em JHEP} {\bf 09} (2020) 085,
  [\href{http://arxiv.org/abs/2003.10265}{{\tt arXiv:2003.10265}}].

\bibitem{Frandsen:2023vhu}
M.~T. Frandsen, M.~Heikinheimo, M.~Rosenlyst, M.~E. Thing, and K.~Tuominen {\em
  JHEP} {\bf 09} (2023) 022, [\href{http://arxiv.org/abs/2302.09104}{{\tt
  arXiv:2302.09104}}].

\bibitem{Reichert:2022naa}
M.~Reichert and Z.-W. Wang {\em EPJ Web Conf.} {\bf 274} (2022) 08003,
  [\href{http://arxiv.org/abs/2211.08877}{{\tt arXiv:2211.08877}}].

\bibitem{Fujikura:2023fbi}
K.~Fujikura, Y.~Nakai, R.~Sato, and Y.~Wang {\em JHEP} {\bf 09} (2023) 053,
  [\href{http://arxiv.org/abs/2306.01305}{{\tt arXiv:2306.01305}}].

\bibitem{Delaunay:2007wb}
C.~Delaunay, C.~Grojean, and J.~D. Wells {\em JHEP} {\bf 04} (2008) 029,
  [\href{http://arxiv.org/abs/0711.2511}{{\tt arXiv:0711.2511}}].

\bibitem{Kurup:2017dzf}
G.~Kurup and M.~Perelstein {\em Phys. Rev. D} {\bf 96} (2017), no.~1 015036,
  [\href{http://arxiv.org/abs/1704.03381}{{\tt arXiv:1704.03381}}].

\bibitem{VonHarling:2017yew}
B.~von Harling and G.~Servant {\em JHEP} {\bf 01} (2018) 159,
  [\href{http://arxiv.org/abs/1711.11554}{{\tt arXiv:1711.11554}}].

\bibitem{Azatov:2019png}
A.~Azatov, D.~Barducci, and F.~Sgarlata {\em JCAP} {\bf 07} (2020) 027,
  [\href{http://arxiv.org/abs/1910.01124}{{\tt arXiv:1910.01124}}].

\bibitem{Ghosh:2020ipy}
T.~Ghosh, H.-K. Guo, T.~Han, and H.~Liu {\em JHEP} {\bf 07} (2021) 045,
  [\href{http://arxiv.org/abs/2012.09758}{{\tt arXiv:2012.09758}}].

\bibitem{Aoki:2021oez}
M.~Aoki, T.~Komatsu, and H.~Shibuya {\em PTEP} {\bf 2022} (2022), no.~6 063B05,
  [\href{http://arxiv.org/abs/2106.03439}{{\tt arXiv:2106.03439}}].

\bibitem{Badziak:2022ltm}
M.~Badziak and I.~Nalecz {\em JHEP} {\bf 02} (2023) 185,
  [\href{http://arxiv.org/abs/2212.09776}{{\tt arXiv:2212.09776}}].

\bibitem{Blasi:2022woz}
S.~Blasi and A.~Mariotti {\em Phys. Rev. Lett.} {\bf 129} (2022), no.~26
  261303, [\href{http://arxiv.org/abs/2203.16450}{{\tt arXiv:2203.16450}}].

\bibitem{Agrawal:2023cgp}
P.~Agrawal, S.~Blasi, A.~Mariotti, and M.~Nee
  \href{http://arxiv.org/abs/2312.06749}{{\tt arXiv:2312.06749}}.

\bibitem{Banerjee:2024qiu}
U.~Banerjee, S.~Chakraborty, S.~Prakash, and S.~U. Rahaman
  \href{http://arxiv.org/abs/2402.02914}{{\tt arXiv:2402.02914}}.

\bibitem{DelleRose:2019pgi}
L.~Delle~Rose, G.~Panico, M.~Redi, and A.~Tesi {\em JHEP} {\bf 04} (2020) 025,
  [\href{http://arxiv.org/abs/1912.06139}{{\tt arXiv:1912.06139}}].

\bibitem{VonHarling:2019gme}
B.~Von~Harling, A.~Pomarol, O.~Pujol\`as, and F.~Rompineve {\em JHEP} {\bf 04}
  (2020) 195, [\href{http://arxiv.org/abs/1912.07587}{{\tt arXiv:1912.07587}}].

\bibitem{Halverson:2020xpg}
J.~Halverson, C.~Long, A.~Maiti, B.~Nelson, and G.~Salinas {\em JHEP} {\bf 05}
  (2021) 154, [\href{http://arxiv.org/abs/2012.04071}{{\tt arXiv:2012.04071}}].

\bibitem{Morgante:2022zvc}
E.~Morgante, N.~Ramberg, and P.~Schwaller {\em Phys. Rev. D} {\bf 107} (2023),
  no.~3 036010, [\href{http://arxiv.org/abs/2210.11821}{{\tt
  arXiv:2210.11821}}].

\bibitem{Jinno:2016knw}
R.~Jinno and M.~Takimoto {\em Phys. Rev. D} {\bf 95} (2017), no.~1 015020,
  [\href{http://arxiv.org/abs/1604.05035}{{\tt arXiv:1604.05035}}].

\bibitem{Addazi:2023ftv}
A.~Addazi, A.~Marcian\`o, A.~P. Morais, R.~Pasechnik, J.~a. Viana, and H.~Yang
  {\em JCAP} {\bf 09} (2023) 026, [\href{http://arxiv.org/abs/2304.02399}{{\tt
  arXiv:2304.02399}}]. [Erratum: JCAP 03, E01 (2024)].

\bibitem{Espinosa:2010hh}
J.~R. Espinosa, T.~Konstandin, J.~M. No, and G.~Servant {\em JCAP} {\bf 1006}
  (2010) 028, [\href{http://arxiv.org/abs/1004.4187}{{\tt arXiv:1004.4187}}].

\bibitem{Giese:2020rtr}
F.~Giese, T.~Konstandin, and J.~van~de Vis {\em JCAP} {\bf 07} (2020), no.~07
  057, [\href{http://arxiv.org/abs/2004.06995}{{\tt arXiv:2004.06995}}].

\bibitem{Giese:2020znk}
F.~Giese, T.~Konstandin, K.~Schmitz, and J.~van~de Vis {\em JCAP} {\bf 01}
  (2021) 072, [\href{http://arxiv.org/abs/2010.09744}{{\tt arXiv:2010.09744}}].

\bibitem{Wang:2021dwl}
X.~Wang, F.~P. Huang, and Y.~Li {\em Phys. Rev. D} {\bf 105} (2022), no.~10
  103513, [\href{http://arxiv.org/abs/2112.14650}{{\tt arXiv:2112.14650}}].

\bibitem{Ajmi:2022nmq}
M.~A. Ajmi and M.~Hindmarsh {\em Phys. Rev. D} {\bf 106} (2022), no.~2 023505,
  [\href{http://arxiv.org/abs/2205.04097}{{\tt arXiv:2205.04097}}].

\bibitem{Tenkanen:2022tly}
T.~V.~I. Tenkanen and J.~van~de Vis {\em JHEP} {\bf 08} (2022) 302,
  [\href{http://arxiv.org/abs/2206.01130}{{\tt arXiv:2206.01130}}].

\bibitem{Wang:2022lyd}
S.-J. Wang and Z.-Y. Yuwen {\em JCAP} {\bf 10} (2022) 047,
  [\href{http://arxiv.org/abs/2206.01148}{{\tt arXiv:2206.01148}}].

\bibitem{Wang:2023jto}
X.~Wang, C.~Tian, and F.~P. Huang {\em JCAP} {\bf 07} (2023) 006,
  [\href{http://arxiv.org/abs/2301.12328}{{\tt arXiv:2301.12328}}].

\bibitem{Caprini:2019egz}
C.~Caprini et~al. \href{http://arxiv.org/abs/1910.13125}{{\tt
  arXiv:1910.13125}}.

\bibitem{Hindmarsh:2015qta}
M.~Hindmarsh, S.~J. Huber, K.~Rummukainen, and D.~J. Weir {\em Phys. Rev. D}
  {\bf 92} (2015), no.~12 123009, [\href{http://arxiv.org/abs/1504.03291}{{\tt
  arXiv:1504.03291}}].

\bibitem{Hindmarsh:2017gnf}
M.~Hindmarsh, S.~J. Huber, K.~Rummukainen, and D.~J. Weir {\em Phys. Rev.} {\bf
  D96} (2017), no.~10 103520, [\href{http://arxiv.org/abs/1704.05871}{{\tt
  arXiv:1704.05871}}].

\bibitem{Athron:2023xlk}
P.~Athron, C.~Bal\'azs, A.~Fowlie, L.~Morris, and L.~Wu {\em Prog. Part. Nucl.
  Phys.} {\bf 135} (2024) 104094, [\href{http://arxiv.org/abs/2305.02357}{{\tt
  arXiv:2305.02357}}].

\bibitem{Ignatius:1993qn}
J.~Ignatius, K.~Kajantie, H.~Kurki-Suonio, and M.~Laine {\em Phys. Rev. D} {\bf
  49} (1994) 3854--3868, [\href{http://arxiv.org/abs/astro-ph/9309059}{{\tt
  astro-ph/9309059}}].

\bibitem{Laine:1993ey}
M.~Laine {\em Phys. Rev. D} {\bf 49} (1994) 3847--3853,
  [\href{http://arxiv.org/abs/hep-ph/9309242}{{\tt hep-ph/9309242}}].

\bibitem{Kurki-Suonio:1995rrv}
H.~Kurki-Suonio and M.~Laine {\em Phys. Rev. D} {\bf 51} (1995) 5431--5437,
  [\href{http://arxiv.org/abs/hep-ph/9501216}{{\tt hep-ph/9501216}}].

\bibitem{Giombi:2023jqq}
L.~Giombi and M.~Hindmarsh {\em JCAP} {\bf 03} (2024) 059,
  [\href{http://arxiv.org/abs/2307.12080}{{\tt arXiv:2307.12080}}].

\bibitem{PhysRevD.106.103524}
D.~Cutting, E.~Vilhonen, and D.~J. Weir {\em Phys. Rev. D} {\bf 106} (Nov,
  2022) 103524.

\bibitem{Lewicki:2023mik}
M.~Lewicki, K.~M\"u\"ursepp, J.~Pata, M.~Vasar, V.~Vaskonen, and H.~Veerm\"ae
  {\em Phys. Rev. D} {\bf 108} (2023), no.~3 036023,
  [\href{http://arxiv.org/abs/2305.07702}{{\tt arXiv:2305.07702}}].

\bibitem{Buen-Abad:2023hex}
M.~A. Buen-Abad, J.~H. Chang, and A.~Hook {\em Phys. Rev. D} {\bf 108} (2023),
  no.~3 036006, [\href{http://arxiv.org/abs/2305.09712}{{\tt
  arXiv:2305.09712}}].

\bibitem{Kolesova:2023mno}
H.~Kolesova and M.~Laine {\em Phys. Lett. B} {\bf 851} (2024) 138553,
  [\href{http://arxiv.org/abs/2311.03718}{{\tt arXiv:2311.03718}}].

\bibitem{Casalderrey-Solana:2022rrn}
J.~Casalderrey-Solana, D.~Mateos, and M.~Sanchez-Garitaonandia
  \href{http://arxiv.org/abs/2210.03171}{{\tt arXiv:2210.03171}}.

\bibitem{Caprini:2011uz}
C.~Caprini and J.~M. No {\em JCAP} {\bf 01} (2012) 031,
  [\href{http://arxiv.org/abs/1111.1726}{{\tt arXiv:1111.1726}}].

\bibitem{RezBook}
L.~Rezzolla and O.~Zanotti, {\em {Relativistic Hydrodynamics}}.
\newblock Oxford University Press, 09, 2013.

\bibitem{Giulini2015LucianoRA}
D.~Giulini {\em General Relativity and Gravitation} {\bf 47} (2015) 1--2.

\bibitem{bookLandau}
L.~Landau and E.~Lifshitz, {\em Fluid Mechanics: Landau and Lifshitz: Course of
  Theoretical Physics}.
\newblock 09, 2013.

\bibitem{Laine:1994bf}
M.~Laine {\em Phys. Lett. B} {\bf 335} (1994) 173--178,
  [\href{http://arxiv.org/abs/hep-ph/9406268}{{\tt hep-ph/9406268}}].

\bibitem{Laine:1998jb}
M.~Laine and K.~Rummukainen {\em Nucl. Phys. B Proc. Suppl.} {\bf 73} (1999)
  180--185, [\href{http://arxiv.org/abs/hep-lat/9809045}{{\tt
  hep-lat/9809045}}].

\bibitem{Cutting:2022zgd}
D.~Cutting, E.~Vilhonen, and D.~J. Weir {\em Phys. Rev. D} {\bf 106} (2022),
  no.~10 103524, [\href{http://arxiv.org/abs/2204.03396}{{\tt
  arXiv:2204.03396}}].

\bibitem{Heinz1938}
L.~Heinz {\em NTRS - NASA Technical Reports Server} {\bf 19930094517} (Jun,
  1939).

\bibitem{Moore:1995si}
G.~D. Moore and T.~Prokopec {\em Phys. Rev.} {\bf D52} (1995) 7182--7204,
  [\href{http://arxiv.org/abs/hep-ph/9506475}{{\tt hep-ph/9506475}}].

\bibitem{Moore:1995ua}
G.~D. Moore and T.~Prokopec {\em Phys. Rev. Lett.} {\bf 75} (1995) 777--780,
  [\href{http://arxiv.org/abs/hep-ph/9503296}{{\tt hep-ph/9503296}}].

\bibitem{Ai:2024shx}
W.-Y. Ai, X.~Nagels, and M.~Vanvlasselaer {\em JCAP} {\bf 03} (2024) 037,
  [\href{http://arxiv.org/abs/2401.05911}{{\tt arXiv:2401.05911}}].

\bibitem{Bodeker:2009qy}
D.~Bodeker and G.~D. Moore {\em JCAP} {\bf 0905} (2009) 009,
  [\href{http://arxiv.org/abs/0903.4099}{{\tt arXiv:0903.4099}}].

\bibitem{Ai:2021kak}
W.-Y. Ai, B.~Garbrecht, and C.~Tamarit {\em JCAP} {\bf 03} (2022), no.~03 015,
  [\href{http://arxiv.org/abs/2109.13710}{{\tt arXiv:2109.13710}}].

\bibitem{Ai:2023see}
W.-Y. Ai, B.~Laurent, and J.~van~de Vis {\em JCAP} {\bf 07} (2023) 002,
  [\href{http://arxiv.org/abs/2303.10171}{{\tt arXiv:2303.10171}}].

\bibitem{Konstandin:2010dm}
T.~Konstandin and J.~M. No {\em JCAP} {\bf 02} (2011) 008,
  [\href{http://arxiv.org/abs/1011.3735}{{\tt arXiv:1011.3735}}].

\bibitem{Balaji:2020yrx}
S.~Balaji, M.~Spannowsky, and C.~Tamarit {\em JCAP} {\bf 03} (2021) 051,
  [\href{http://arxiv.org/abs/2010.08013}{{\tt arXiv:2010.08013}}].

\bibitem{Sanchez-Garitaonandia:2023zqz}
M.~Sanchez-Garitaonandia and J.~van~de Vis
  \href{http://arxiv.org/abs/2312.09964}{{\tt arXiv:2312.09964}}.

\bibitem{Krajewski:2024gma}
T.~Krajewski, M.~Lewicki, and M.~Zych {\em JHEP} {\bf 05} (2024) 011,
  [\href{http://arxiv.org/abs/2402.15408}{{\tt arXiv:2402.15408}}].

\bibitem{Laurent:2022jrs}
B.~Laurent and J.~M. Cline {\em Phys. Rev. D} {\bf 106} (2022), no.~2 023501,
  [\href{http://arxiv.org/abs/2204.13120}{{\tt arXiv:2204.13120}}].

\bibitem{Laurent:2020gpg}
B.~Laurent and J.~M. Cline {\em Phys. Rev. D} {\bf 102} (2020), no.~6 063516,
  [\href{http://arxiv.org/abs/2007.10935}{{\tt arXiv:2007.10935}}].

\bibitem{DeCurtis:2022hlx}
S.~De~Curtis, L.~D. Rose, A.~Guiggiani, A.~G. Muyor, and G.~Panico
  \href{http://arxiv.org/abs/2201.08220}{{\tt arXiv:2201.08220}}.

\bibitem{DeCurtis:2023hil}
S.~De~Curtis, L.~Delle~Rose, A.~Guiggiani, A.~Gil~Muyor, and G.~Panico {\em
  JHEP} {\bf 05} (2023) 194, [\href{http://arxiv.org/abs/2303.05846}{{\tt
  arXiv:2303.05846}}].

\bibitem{DeCurtis:2024hvh}
S.~De~Curtis, L.~Delle~Rose, A.~Guiggiani, A.~Gil~Muyor, and G.~Panico {\em
  JHEP} {\bf 05} (2024) 009, [\href{http://arxiv.org/abs/2401.13522}{{\tt
  arXiv:2401.13522}}].

\bibitem{Dine:1992wr}
M.~Dine, R.~G. Leigh, P.~Y. Huet, A.~D. Linde, and D.~A. Linde {\em Phys. Rev.}
  {\bf D46} (1992) 550--571, [\href{http://arxiv.org/abs/hep-ph/9203203}{{\tt
  hep-ph/9203203}}].

\bibitem{Mancha:2020fzw}
M.~Barroso~Mancha, T.~Prokopec, and B.~Swiezewska
  \href{http://arxiv.org/abs/2005.10875}{{\tt arXiv:2005.10875}}.

\bibitem{Bodeker:2017cim}
D.~Bodeker and G.~D. Moore {\em JCAP} {\bf 1705} (2017), no.~05 025,
  [\href{http://arxiv.org/abs/1703.08215}{{\tt arXiv:1703.08215}}].

\bibitem{Vanvlasselaer:2020niz}
A.~Azatov and M.~Vanvlasselaer {\em JCAP} {\bf 01} (2021) 058,
  [\href{http://arxiv.org/abs/2010.02590}{{\tt arXiv:2010.02590}}].

\bibitem{Gouttenoire:2021kjv}
Y.~Gouttenoire, R.~Jinno, and F.~Sala {\em JHEP} {\bf 05} (2022) 004,
  [\href{http://arxiv.org/abs/2112.07686}{{\tt arXiv:2112.07686}}].

\bibitem{Azatov:2023xem}
A.~Azatov, G.~Barni, R.~Petrossian-Byrne, and M.~Vanvlasselaer {\em JHEP} {\bf
  05} (2024) 294, [\href{http://arxiv.org/abs/2310.06972}{{\tt
  arXiv:2310.06972}}].

\bibitem{Ai:2023suz}
W.-Y. Ai {\em JCAP} {\bf 10} (2023) 052,
  [\href{http://arxiv.org/abs/2308.10679}{{\tt arXiv:2308.10679}}].

\bibitem{Azatov:2024auq}
A.~Azatov, G.~Barni, and R.~Petrossian-Byrne
  \href{http://arxiv.org/abs/2405.19447}{{\tt arXiv:2405.19447}}.

\bibitem{Megevand:2014yua}
A.~Megevand and F.~A. Membiela {\em Phys. Rev. D} {\bf 89} (2014), no.~10
  103503, [\href{http://arxiv.org/abs/1402.5791}{{\tt arXiv:1402.5791}}].

\bibitem{Megevand:2013yua}
A.~Megevand and F.~A. Membiela {\em Phys. Rev. D} {\bf 89} (2014), no.~10
  103507, [\href{http://arxiv.org/abs/1311.2453}{{\tt arXiv:1311.2453}}].

\bibitem{PhysRev.94.1468}
A.~H. Taub {\em Phys. Rev.} {\bf 94} (Jun, 1954) 1468--1470.

\bibitem{1973ApJ...179..897T}
K.~S. {Thorne} {\em Astrophysical Journal} {\bf 179} (Feb., 1973) 897--908.

\end{thebibliography}\endgroup

\end{document}